\documentclass[twocolumn]{aastex62}
\usepackage{natbib}

\newcommand{\PaulsCode}{\texttt{RT1D}}
\newcommand{\Hal}{\mathrm{H}\alpha}

\newcommand{\Msun}{M_\odot}
\newcommand{\gamad}{\gamma_\mathrm{ad}}

\newcommand{\days}{{\rm days}}
\newcommand{\cm}{\mathrm{cm}}
\newcommand{\kms }{{\rm km~s^{-1}}} 
\newcommand{\densu}{{\rm g~cm^{ -3}}} 
\newcommand{\timp}{t_{\rm imp}} 
\newcommand{\tx}{t_{\rm xwall}} 
\newcommand{\vejmax}{v_\mathrm{ej, max}}
\newcommand{\vshock}{v_\mathrm{sh}}
\newcommand{\uej}{u_\mathrm{ej}}
\newcommand{\rhoej}{\rho_{\rm ej}}
\newcommand{\rhocsm}{\rho_{\rm csm}}
\newcommand{\rhowall}{\rho_\mathrm{wall}}

\newcommand{\Rsw}{R_\mathrm{c,0}}

\newcommand{\Rfwd}{R_\mathrm{fwd}}
\newcommand{\Rrev}{R_\mathrm{rev}}
\newcommand{\Rwall}{R_\mathrm{wall}}
\newcommand{\wall}{{\rm wall}}
\shorttitle{}
\shortauthors{Harris \& Nugent}

\begin{document}


  \title{Outside the Wall: Hydrodynamics of Type I Supernovae Interacting with 
         a Partially Swept-Up Circumstellar Medium}

\author[0000-0002-1751-7474]{C.\,E.\,Harris}
\affiliation{Department of Physics and Astronomy, Michigan State University, East Lansing, MI 48824, USA }
\author[[0000-0002-3389-0586]{P.\,E.\,Nugent}
\affiliation{Lawrence Berkeley National Laboratory, 1 Cyclotron Road, MS 50B-4206, Berkeley, CA 94720, USA}
\affiliation{Department of Astronomy, University of California, Berkeley, CA 94720-3411, USA}

\correspondingauthor{CEH}
\email{harr1561@msu.edu}


\begin{abstract}    
    Explaining the observed diversity of supernovae (SNe) and the 
    physics of explosion requires knowledge of their progenitor stars,
    which can be obtained by constraining the circumstellar medium (CSM).
    Models of the SN ejecta colliding with CSM are
    necessary to infer the structure of the CSM and tie it back
    to a progenitor model.
    Recent SNe\,I revealed CSM concentrated at a distance 
    $r \sim10^{16}~\cm$, 
    for which models of SN interaction are extremely limited.
    In this paper, we assume the concentrated region is a ``wall''
    representing swept-up material, and unswept material lies outside
    the wall.
    We simulate one-dimensional hydrodynamics
    of SNe\,Ia~\&~Ib impacting $~300$ unique CSM configurations
    using \PaulsCode, which captures the Rayleigh-Taylor instability.
    We find that the density ratio between the wall and ejecta --
    denoted $A_0$ or ``wall height'' -- 
    is key, and higher walls deviate more from self-similar evolution.
    Functional fits accounting for $A_0$ are presented
    for the forward shock radius evolution.
    We show that higher walls have more degeneracy between CSM properties
    in the deceleration parameter, 
    slower shocks, deeper-probing reverse shocks, slower shocked
    ejecta, less ejecta mass than CSM in the shock, 
    and more mixing of ejecta into the CSM at early times.
    We analyze observations of SN\,2014C (Type\,Ib) and suggest
    that it had a moderately high wall ($10 \lesssim A_0\lesssim200$)
    and wind-like outer CSM.
    We also postulate an alternate interpretation for the 
    radio data of SN\,2014C, that the radio rise occurs in the wind 
    rather than the wall.
    Finally, we find that hydrodynamic measurements at very 
    late times cannot distinguish the presence of a wall, 
    except perhaps as an anomalously wide shock region.
\end{abstract}



\section{Introduction}\label{sec:intro}


Supernovae (SNe) infuse their host galaxies with metals and 
energy \citep{Tinsley1980},
accelerate particles \citep{BlandfordOstriker78},
create compact objects \citep{1968Sci...162.1481S, 1969Natur.221..525C},  
and give us a way of measuring cosmic expansion \citep{Wagoner77}.
Using them precisely and accurately for these purposes requires 
detailed knowledge of their stellar progenitors.
In principle the progenitors can be directly identified from 
pre-explosion observations \citep{2009ARA&A..47...63S}, however, most systems 
are too dim to do this \citep{Bloom+12}. 
Therefore, progenitors are typically constrained through circumstantial
evidence that can be connected to theoretical models.

The circumstellar medim (CSM) fossilizes 
stellar evolution through the millennia before explosion,
and is thus diagnostic of mass-loss and mass-transfer processes
that are central to SN progenitor identification.
The CSM is illuminated by the blast wave formed when the SN ejecta impact it.
Using models of the shock propagation, this light can be translated into the
structure of the CSM and ejecta.

Normal luminosity (i.e., not superluminous) Type\,I SNe are the focus of this work.
SNe\,I lack hydrogen in their spectra because the progenitor lost its
hydrogen envelope.
The two groups of normal-luminosity SNe\,I are SNe\,Ia (thermonuclear) 
and SNe\,Ibc (core-collapse). 
SNe\,Ia occur in binary star systems, and the primary mystery of their
progenitors is the nature of the mass-donor companion star to the 
carbon-oxygen white dwarf that explodes \citep[e.g.,][]{2014ARA&A..52..107M}.
The CSM of SNe\,Ia gives us insight into the mass transfer process 
and nature of the companion \citep[e.g.,][]{Chomiuk+12}.
SNe\,Ibc may not occur in binary systems, although they likely do \citep[e.g.,][]{2010ApJ...721..777A}.
The CSM of SNe\,Ibc tells us about the timescale and physical mechanism
by which the outer envelope is lost \citep[e.g.,][]{Weiler+02}.

Discovered in increasing numbers and across all SN classes 
are SNe with dense CSM at $\gtrsim10^{16}~\cm$, and an evacuated
cavity within this distance.
Unlike in the canonical interaction scenario that produces SNe\,IIn,
the SN initially appears normal (non-interacting) while it traverses the
evacuated cavity.
Months after explosion, ejecta impact dense CSM
and transform into an interacting SN. 
For this reason we refer to such scenarios as ``delayed interaction.''
We use the shorthand ``SNe\,X;n'' to refer to delayed-interaction events, 
where ``X'' gets replaced by the non-interacting/peak-light/original
classification, ``n'' refers to narrow emission lines as for SNe\,IIn, 
and the semicolon represents the separation between 
non-interacting and interacting phases.
The most famous supernova SN\,1987A is an example of an SN\,II;n. 
SN\,2014C is an SN\,Ib;n -- albeit with some (perhaps $\sim0.03~\Msun$) hydrogen
left in its outer evelope \citep{Milisav+15} reminiscent of SNe\,IIb --
whose interaction began around 100 days after explosion.
SN\,2014C has a remarkable dataset, with high-cadence radio observations 
throughout its evolution complemented by x-ray, optical, and infrared
spectra.
Examples of SNe\,Ia;n are SN\,2002ic \citep{WoodVasey04}, 
PTF\,11kx \citep{Dilday+12,Silverman+13b,GH+17}, and 
SN\,2015cp \citep{Graham+19}.

    There are two enormous gifts in studying SNe\,I;n.
    First, that hydrogen lines in the spectra are clean tracers of the 
    CSM. 
    Second, that the non-interacting portion of the light-curve can be 
    analyzed with existing light-curve analysis tools to derive the SN 
    properties. 
    This means the SN structure can be constrained in the shock modeling
    much better than is usually the case, breaking ejecta-CSM degeneracies.


    Understanding the physical origin of CSM in SNe\,I;n whose
    dense ($\sim10^{-18}~\densu$) CSM appears to lie at $\sim10^{16}~\cm$ is a challenge.
    Such an environment can be created by a sudden increase 
    in the ram pressure of the mass being expelled by the star system.
    A nova eruption from the system 
    is an example of such a process \citep{MooreBildsten12,Dimitriadis+14}.
    Another example is the switch from a slow wind to a fast one
    \citep[e.g.,][]{Castor+75,Weaver+77,RamirezRuiz+05}.
    The eruptive models naturally build up mass at $\sim10^{16-17}~\cm$, 
    while models of ``wind-blown bubbles'' generally place the over-density
    at a larger distance ($\sim10^{18-19}~\cm$). 
    One challenge for both models is to explain the high mass
    ($\sim0.1~\Msun$) 
    of material that is observed to exist at these distances.
    Thus, while both eruption and wind avenues are broadly successful
    for creating detached CSM, they fail in detail.
    Common to both is a CSM structure shaped by the sweeping-up of
    an existing medium.

    For SNe\,Ib, the wind-blown bubble scenario has been modeled
    throughout the decades (owing to their proposed relationship
    to Wolf-Rayet stars).
    Typically these models have applicability to SN remnants, 
    since the walls are not impacted until
    decades or centuries post-explosion.
    \citet{ChevalierLiang89} present analytic estimates for the
    evolution of the shock while it is in the wall. 
    \citet{Dwarkadas05} explores one-dimensional numerical calculations
    of the formation of the wall and evolution of the SN through the
    cavity, wall, and outer medium.
    Although wind-blown bubbles are supposed to form at large radii,
    the basic CSM structure has been compacted down for bespoke
    models of specific SNe, e.g. for the SNe\,Ib;n 2001em and 1996cr 
    \citep{ChugaiChevalier06,Dwarkadas+10}.
    Even for this specific CSM formation scenario, a systematic
    numerical study resulting in quantitative relationships that can
    be applied to new SNe has not been carried out.

    Here we study SNe\,Ia\&Ib impacting CSM that has been 
    partially swept into a ``wall,'' 
    outside of which lies the original pre-swept medium,
    as in wind-blown bubbles or nova eruptions.
    Figure~\ref{fig:init_conds} illustrates this configuration.
    Our aim is to provide interpretation tools for future SNe or ensemble datasets.
    In this paper, we limit our scope to the hydrodynamic evolution of the shocks,
    leaving the radiation calculation to a sequel. 
    Some observations directly probe the kinematic properties of the shock fronts, 
    e.g., spectroscopic line profiles or
    very long baseline radio interferometry (VLBI). 
    Most of these observations are interpreted in the self-similar/``mini-shell''
    framework of \citet{Chevalier82a} (hereafter ``C82'') 
    which has been developed in many papers over the decades.
    Therefore we particularly compare the hydrodynamic evolution of our models
    to what one would derive from the mini-shell model given the SN and CSM
    properties.
    To summarize some of the important limitations of the mini-shell model:
    it is a solution only applicable while the reverse shock is in the 
    outermost ejecta, assumes the ejecta and CSM are similar density, 
    and does not apply to transient phases when the forward shock crosses into 
    a new CSM profile (e.g., after crossing the wall).


    Our models are the first to systematically explore with numerical simulations
    a many-pronged space of CSM properties through a suite of $\sim600$ unique simulations.
    This is also the first suite of one-dimensional interaction simulations to capture the
    Rayleigh-Taylor instability, using \PaulsCode\ \citep{Duffell16} -- 
    though, of course, only approximately.
    The Rayleigh-Taylor instability prevents artificial density discontinuities
    from occurring in nature; in numerical simulations, these discontinuities
    can complicate the analysis of simulations when working with a complex CSM 
    structure \citep[e.g.,][]{ChevalierLiang89,Dwarkadas05}. 
    Throughout this paper, we highlight the SN\,Ib models
    because of the emphasis on wind-blown bubbles in the literature
    and the quality of the SN\,2014C data.

    The paper is organized as follows.
    In \S\,\ref{sec:models} we describe the free parameters of our model suite
    (see Figure~\ref{fig:init_conds}) and the hydrodynamics code used
    to simulate the interaction.
    In \S\ref{sec:analysis} we present the properties of the shock fronts and
    shocked gas in our model suite, focusing on quantities frequently
    used in interpreting observations: shock radii, shock front speeds,
    ejecta deceleration, shocked mass, and amount of mixing between
    ejecta and CSM.
    We particularly focus on finding analytic relations to describe
    the evolution of these properties, comparing them to self-similar
    evolution, and looking for degeneracies between CSM parameters.
    In \S\,\ref{sec:disc} we discuss the 
    application of these results to observations in more detail, 
    first considering SN\,2014C VLBI observations to constrain the 
    wall height and outer medium density profile, 
    then reinterpreting SN\,2014C radio observations that
    probe the wall extent,
    and concluding with comments on which hydrodynamic quantities 
    are reliable records of the wall in VLBI observations.
    We leave a detailed comparison between this model suite and
    observations of delayed-interaction SNe\,I to the sequel paper on
    radiation signatures. 
    In \S\ref{sec:conc} we provide a summary.

\section{Simulation Description}\label{sec:models}

    \subsection{Initial Conditions}\label{ssec:mod_setup}

\begin{figure}[h]
        \centering
        \includegraphics[width=\linewidth]{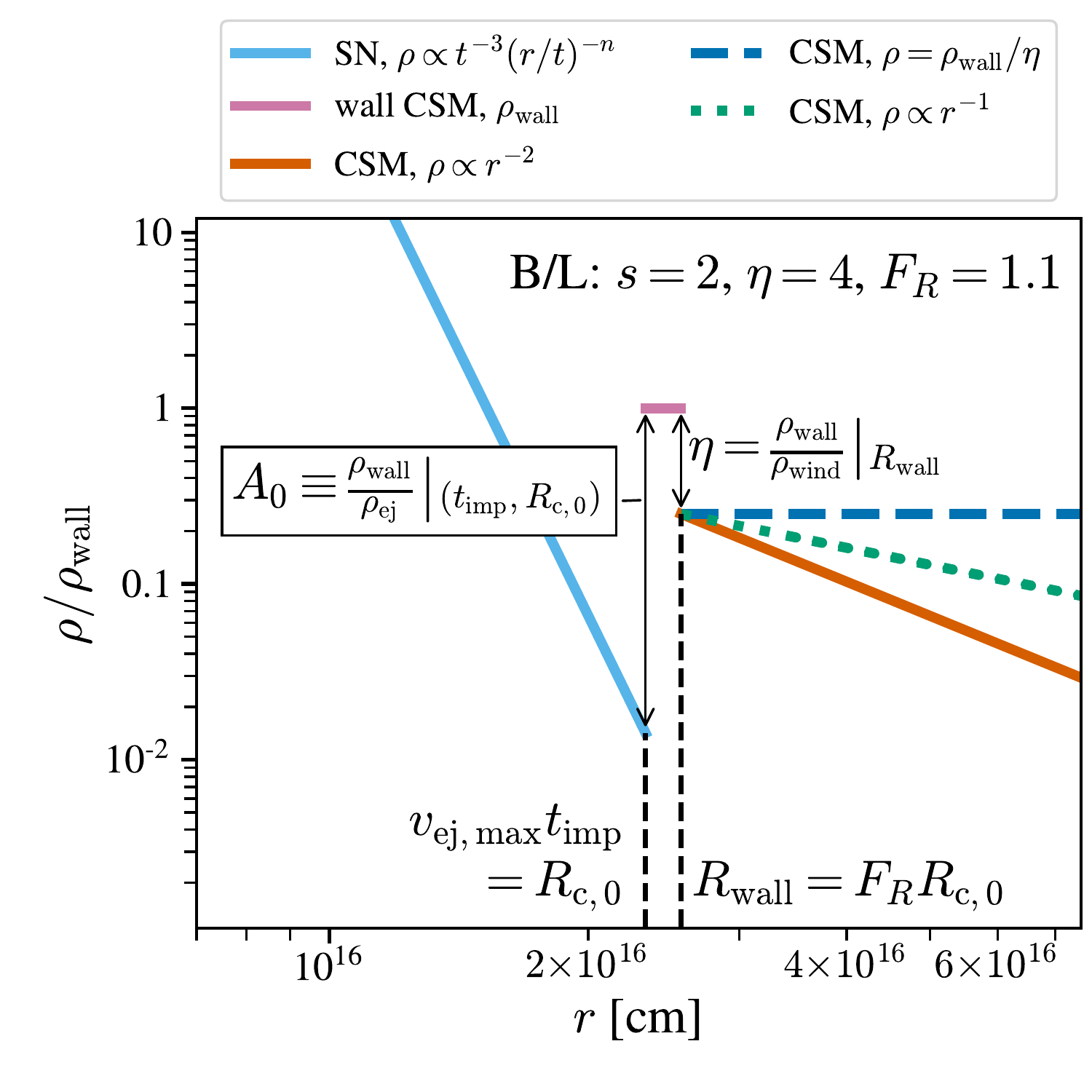}
        \caption{Definition of model variables and illustration of 
        initial conditions. 
        The baseline (B/L) CSM parameters are stated.
        }
        \label{fig:init_conds}
    \end{figure}

    In this paper we investigate the interaction of freely expanding 
    supernova ejecta with a circumstellar medium
    characterized by an innermost evacuated cavity
    terminating at a thin ``wall'' of material, outside of which is 
    an extended, lower density medium that we call the ``outer CSM.''
    Figure~\ref{fig:init_conds} illustrates the model parameters
    detailed in this section.

    We ignore the evolution in the inner evacuated cavity, assuming
    that the low-density medium has a negligible effect on the wall
    and ejecta profiles.

    For the ejecta, we assume free expansion $v=r/t$ 
    ($v$ is velocity, $r$ radius, and $t$ time)
    and consider density profiles that approximate
    an SN\,Ia and an SN\,Ib with a broken power law.
    The broken power-law profile arises from the propagation of
    the explosion shock through the star, which itself has an approximately
    broken power-law structure, as derived in \citet{ChevalierSoker89}.
    The outer regions have $\rhoej = g^n t^{n-3} r^{-n} \propto t^{-3} v^{-n}$.
    The inner regions have a flatter density profile, which we model
    as $\rhoej \propto r^{-1}$. 
    The transition between inner and outer ejecta occurs at the 
    transition velocity $v_t$. 
    Density normalization factors and $v_t$ are calculated
    using the expressions in \citet{Kasen10}.
    For the SN\,Ia, we use $n=10$,
    ejecta mass $M_\mathrm{Ia} = 1.38~\Msun$, 
    and kinetic energy $E_\mathrm{ej}=10^{51}~\mathrm{erg}$ \citep{HNK16}. 
    The SN\,Ib has $n=9$, 
    $M_\mathrm{ej} = 1.7\Msun$, and $E_\mathrm{ej}=1.8\times10^{51}~\mathrm{erg}$ 
    to be consistent with SN\,2014C \citep{Margutti+17}.
    We truncate the ejecta at $\vejmax=30,000~\kms$, and simulations begin 
    when the outermost ejecta reach the inner radius of the wall, $\Rsw$,
    at time $\timp = \Rsw/\vejmax$. 
    We explore models with $\timp = \{30, 60, 90, 120\}~\days$, which all 
    have $\Rsw \sim 10^{16}~\cm$.

    The wall has a constant density
    $\rhowall = \{ 10^{-18}, 3\times10^{-18}, 10^{-17} \}~\densu$, 
    and the outer CSM has a density profile 
    $\rhocsm = q r^{-s}$ with s=\{0, 1, 2\}.
    The outer edge of the wall is located at $\Rwall \equiv F_R \Rsw$
    with $F_R = {1.01,1.03,1.1,1.3}$. 
    We take the density of the wall as proportional to the density of outer
    CSM at $\Rwall$, as would be appropriate for a wall formed by a shock.
    The constant of proportionality is the compression ratio, $\eta$, 
    and is given by the Rankine-Hugoniot strong shock jump conditions
    as 
    \begin{equation}
        \eta \equiv \frac{\rho_\mathrm{shocked}}{\rho_\mathrm{unshocked}} \equiv 
        \frac{\rhowall}{\rhocsm(\Rwall)} = \frac{\gamad+1}{\gamad-1} \; ,
    \end{equation}
    where $\gamad$ is the adiabatic index.
    We explore compression ratios $\eta = \{4, 7\}$.

\begin{figure}
        \centering
        \includegraphics[width=.7\linewidth]{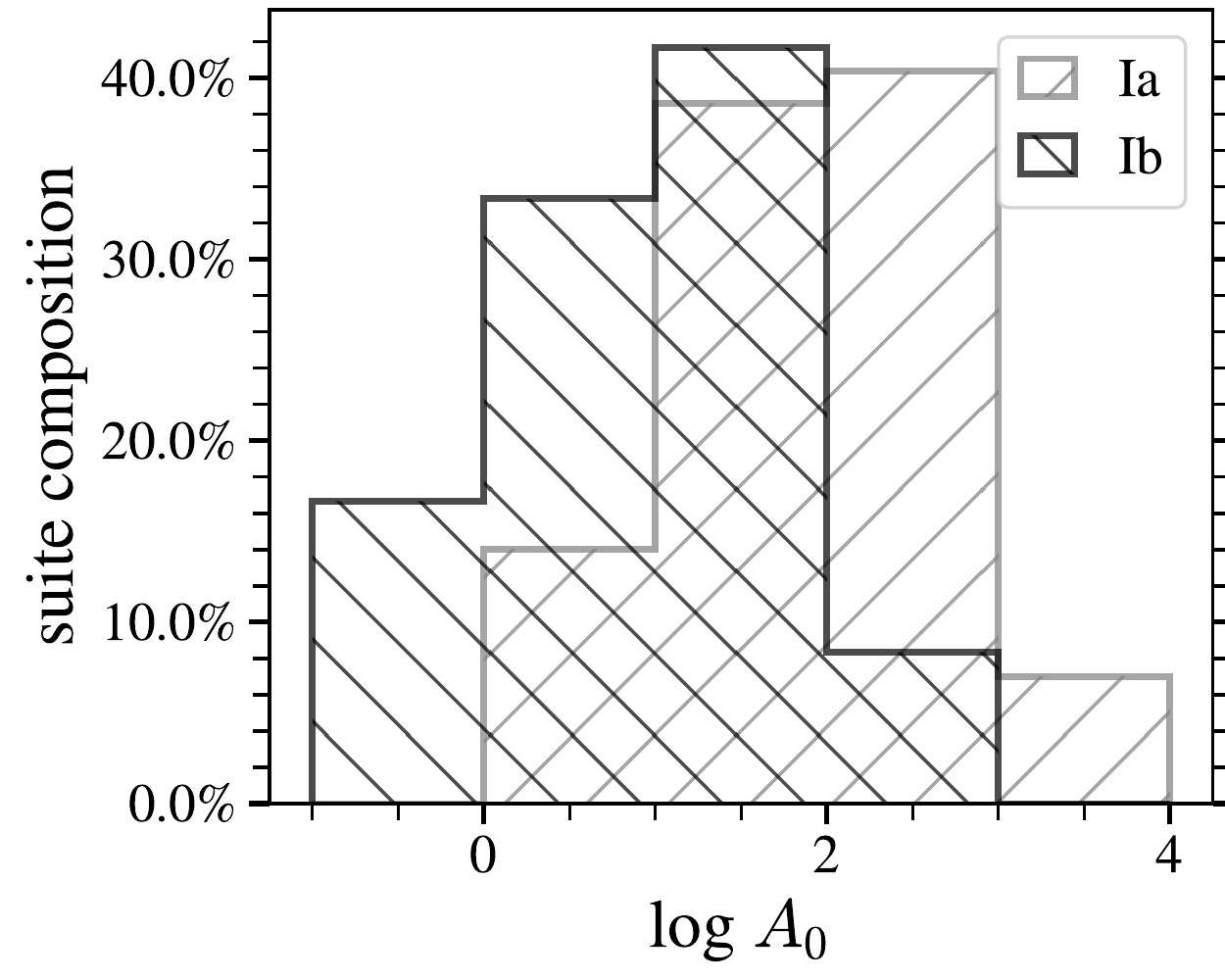}
        \caption{Representation of $A_0$ values across all models,
        divided by SN type.
        Both types have the same CSM suite but the SN\,Ib ejecta
        are denser, shifting $A_0$ to lower values.}
        \label{fig:suite_A0}
    \end{figure}

    The ratio of the CSM  density to the ejecta density at the point
    and time of impact is
    \begin{equation}
        A_0 \equiv \frac{\rhowall}{\rhoej(\Rsw,\timp)} 
            \propto \rhowall\timp^3 
            \propto \frac{M_\wall \timp^3}{F_R^3-1} \; .
        \label{eqn:A0}
    \end{equation}
    The $A_0$ values covered by the suite are shown in Figure~\ref{fig:suite_A0}.
    This parameter is named for its analogy to the $A$ parameter in
    C82, which is algebraically equivalent. 
    The subscript zero is to indicate that it is at the intial time.

    The variation in $\rhowall$, $\timp$, $s$, $\eta$, $F_R$, 
    and SN class create a suite of 576 models.
    Since $A_0$ depends only on $\rhowall$ and $\timp$, each SN has 12 unique $A_0$
    values.
    When it is useful to narrow our focus in a demonstrative graph, we emphasize
    the SN Ib models. 
    We also define the ``baseline'' (B/L) set of CSM parameters to be $s=2$, $\eta=4$, and $F_R=1.1$,
    to parallel the models posited for SN 1996cr and SN 2014C 
    \citep{Dwarkadas+10,Margutti+17}.

    \subsection{Computational Method}\label{ssec:mod_comp}

    We use the \PaulsCode\ code \citep{Duffell16} to model the hydrodynamics of 
    the interaction.
    This is a moving-mesh Eulerian hydrodynamics code with a gamma-law equation
    of state (with $\gamad=5/3$).
    We use the logarithmic spacing option to define the initial grid.
    
    We ensured that the baseline set for both SN types was high enough resolution
    to obtain smooth curves in our analyses and sometimes call these
    ``high-resolution'' baseline models.
    By ``high-resolution'' we mean they were given 8000 zones instead of 
    the 3000 zone limit applied to the rest of the suite.
    The spatial resolution $\Delta x$ is not constant over the grid (logarithmic 
    spacing) or in time (moving mesh).
    These models had significantly longer
    run times than a low-resolution counterpart, hence the restriction
    on non-baseline models, which made the running of the entire suite feasible.
    Depending on the exact gridding, non-baseline models can still have
    comparable resolution to the baseline set.

    The characterizing feature of \PaulsCode\ is that it incorporates a one-dimensional 
    prescription of the Raleigh-Taylor instability based on three-dimensional 
    models. 
    This instability is known to be common in SN interaction.
    Using this code, ejecta are able to mix with CSM, changing the 
    composition of the shocked material, in contrast to the unmixed case. 
    The fraction of CSM is tracked by the passive scalar \texttt{X}.
    
    The gas pressures in our simulations 
    imply very high gas temperatures, leading some to worry about
    the contribution from radiation pressure and accuracy of $\gamad=5/3$.
    However, this concern arises from intuition based on materials 
    radiating as a blackbody, which our low-density gas does not
    (though the gas is mostly in a thermal velocity distribution).
    For blackbody radiation the radiation pressure ($p_\mathrm{rad}$) 
    is proportional to the temperature ($T$) as 
    $p_\mathrm{rad} = aT^4/3$, 
    where $a=7.56\times10^{-15}~\mathrm{erg~cm^{-3}~K^{-4}}$ 
    is the radiation density constant.
    Here instead we must return to the fundamental definition based 
    on energy density,
    $p_\mathrm{rad} = u_\mathrm{rad}/3$, and calculate $u_\mathrm{rad}$
    from the intensity ($I_\nu$) which in the optically thin case is the 
    integration of the emissivity ($j_\nu$) along a path 
    $I_\nu = \int j_\nu ds$. 
    Performing the integrations over frequency and solid angle 
    (assuming isotropic emission) yields 
    $p_\mathrm{rad} = (1/3c) \int \varepsilon_\mathrm{ff} ds 
    \sim \varepsilon_\mathrm{ff} \Delta R_\mathrm{shock} / (3c)$,
    in which $c$ is the speed of light in vacuum,
    $\Delta R_\mathrm{shock}$ is the radial width of the shock,
    and $\varepsilon_\mathrm{ff}$ is the frequency-integrated power of 
    free-free emission. 
    For details of this calculation, we refer the reader to \citet{RybickiLightman1979}.
    In all of these calculations the temperature is the 
    electron temperature which may be lower than the gas (ion) temperature by
    a factor $\lesssim2000$ \citet{Ghavamian+07}.
    For our estimate we will use the ion temperature and therefore overestimate
    the radiation pressure, since $\varepsilon_\mathrm{ff} \propto T^{1/2}$.
    We find that for our models, $p_\mathrm{rad} \lesssim 10^{-4} p_\mathrm{gas}$
    therefore $\gamad=5/3$ applies.
    
    We investigated Bremsstrahlung cooling but found that the timescales
    are too long to be important, in line with \citet{Dwarkadas+10}.


\section{Analysis}\label{sec:analysis}


    In this section we present our analysis of the shock hydrodynamics, 
    focusing on  quantities that have been used 
    to interpret observations of interacting supernovae.
    Our methods for calculating these quantities are described in
    \S~\ref{ssec:an_meth}.

    We find that the evolution depends crucially on the parameter $A_0$, the
    initial density ratio between the CSM wall and the ejecta,
    as did \citet{Dwarkadas05}.
    Keep in mind that $A_0$ is not strictly the density of the wall -- 
    the same density of CSM impacted at a later time will have 
    a higher $A_0$.  
    
\begin{figure}
        \centering
        \includegraphics[width=\linewidth]{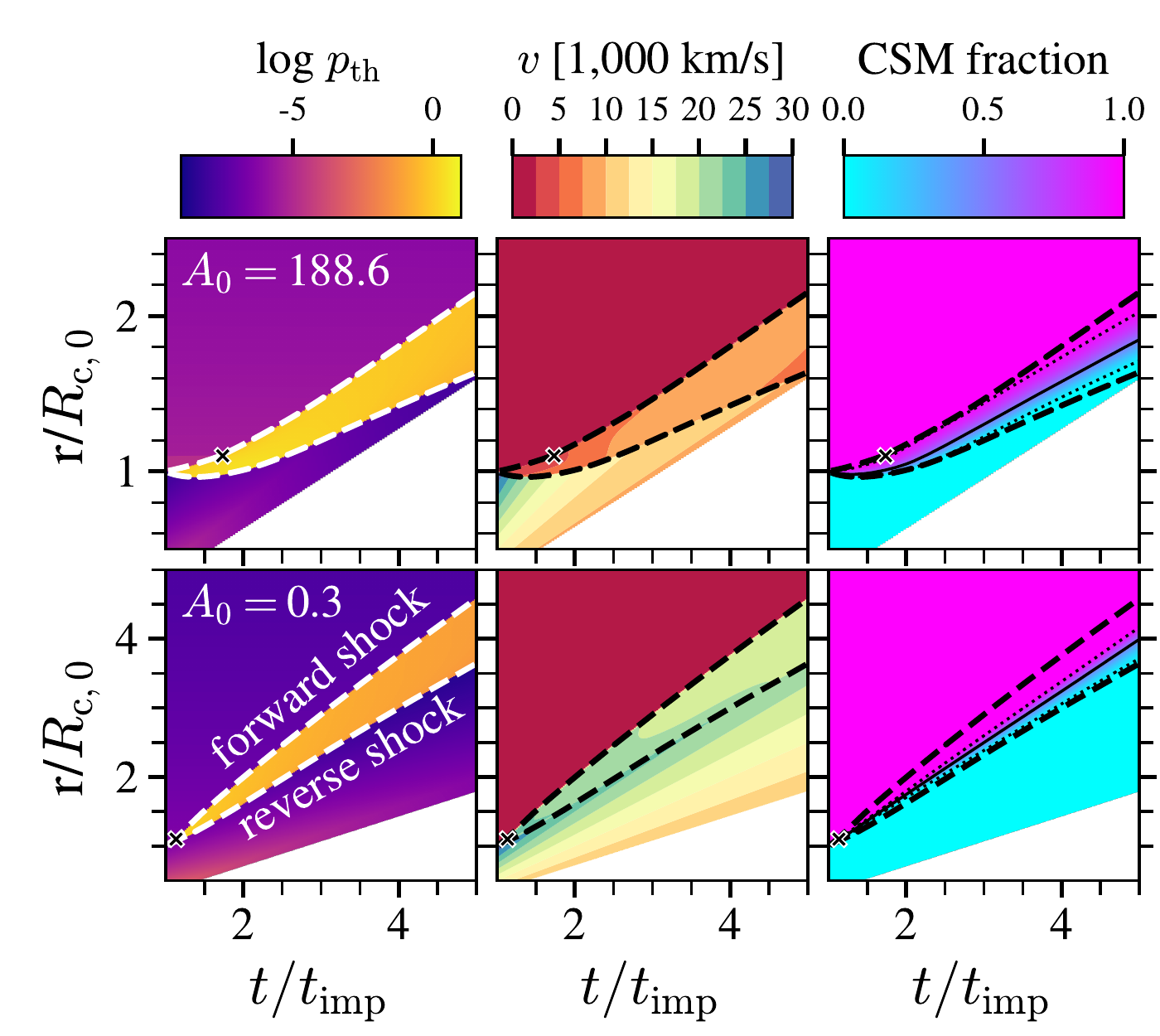}
        \caption{The evolution of a low-$A_0$ model (bottom) and 
                 high-$A_0$ model (top) from the baseline set,
                 in radial coordinates.
                 Note the smaller y-axis range in top panels.
                 Dashed lines show shock fronts, 
                 and ``x'' marks $\tx$.
                 \textit{Left.} The common logarithm of pressure (units $\mathrm{erg~cm^{-3}}$),
                 used to find shock fronts.
                 Notice that the higher wall produces a stronger reverse shock
                 that initially moves inward. 
                 \textit{Middle.} Gas speed. 
                 Notice the lower shocked gas speeds in the higher wall model
                 and that the shock travels deeper into the ejecta, 
                 both effects of the wall's increased stopping power.
                 \textit{Right.} Fraction of gas that is CSM, 
                 with $X=0.5$ (solid) and $X=0.05,0.95$ (dotted) delineated.
                 }
        \label{fig:mod_ev}
    \end{figure}

    Figure~\ref{fig:mod_ev} shows a comparison of a low-$A_0$ (``low wall'') 
    model to one with high-$A_0$ (``high wall'').
    Both models have baseline parameters $s=2$, $\eta=4$, and $F_R=1.1$.
    The x-axis is time normalized to $\timp$, and the y-axis
    is radius normalized to initial contact radius.
    This figure demonstrates many of the themes explored in detail in 
    our analysis.
    
    The left panels show the time evolution of the  pressure, 
    from which the shock fronts are clearly distinguished.
    There is a steep gradient between shocked gas (bright/orange) and 
    pre-shock gas (dark/purple) that is used to identify the shock fronts, 
    which are shown as dashed lines in all panels. 
    The forward shock front moves slower for higher walls -- 
    note that the y-axis range is smaller for the high-wall model.
    Crosses on the line of the forward shock front show the time at which
    the shock crosses the wall ($\tx$), which occurs later for the higher wall. 
    The trade-off is, that the reverse shock is much stronger for the 
    higher walls -- we see that there is actually a period of time when
    the reverse shock is moving inward radially in the higher wall model.
    In short, higher-wall models have a weaker/slower forward shock and stronger
    reverse shock than lower-wall models.
    Analysis of the shock front radii and speeds is presented in
    \S~\ref{ssec:shock_r}-\ref{ssec:shock_v}.
    
    The central panels show the gas velocities.
    Initially, all CSM moves at $100~\kms$ while ejecta are in
    free expansion and have a maximum velocity of $30,000~\kms$. 
    The shocked gas in the higher-wall model is much
    slower than in the lower-wall model.
    High walls have more stopping power than low walls, and we see
    that the reverse shock is reaching farther into the ejecta and 
    probing lower velocities.
    This means the reverse shock of higher-wall models will cross
    into the dense inner ejecta faster than lower-wall models.
    Analysis of ejecta deceleration can be found in \S~\ref{ssec:ej_v}. 
    
    The right panel shows the fraction of gas that is CSM material.
    Because our study uses \PaulsCode, the ejecta and CSM 
    mix due to the Rayleigh-Taylor instability.
    Dotted lines show where the composition is 5\% and 95\% CSM
    to give a sense of the boundary between mixed and unmixed material.
    (In our later analysis of mixing, we use a boundary of 1\% 
    to define mixed material.)
    The amount of mixing depends on the height of the wall and time
    of the simulation.
    Analyses of the amount of mass shocked and amount of mixing can be found in
    \S~\ref{ssec:shock_m} \& \ref{ssec:mixing}.

    \subsection{Methods}\label{ssec:an_meth}
    \paragraph{Identifying Shock Fronts}
    
    We denote the forward and reverse shock radii as $\Rfwd$ (or $R_f$) 
    and $\Rrev$ (or $R_r$),  respectively.
    We identify the forward and reverse shock fronts as the locations of 
    maximum magnitude in the radial gradient of the common logarithm of the 
    thermal pressure, on either side of the ejecta/CSM interface.
    In one-dimensional models without mixing, the ejecta/CSM interface is defined 
    by a contact discontinuity in the mass density. 
    We do not have this discontinuity, and define the interface as 
    the location where the CSM fraction is half.
    
    \PaulsCode\ is particularly suited to capturing shock fronts, and we 
    found $<1\%$ difference in the determination of the forward and reverse
    shock radii with low resolution compared to high resolution.
    For the baseline CSM parameters, we only present high-resolution
    results. 
    
    In some models with low $A_0$ we have observed that the 
    rarefaction wave that propagates into the shocked ejecta (after the
    forward shock crosses from the wall into the lower-density outer CSM)
    steepens into a shock and does not overtake the original reverse shock
    front, creating a double-shock structure at late times.
    Once the rarefaction shock is as steep as the reverse shock, 
    our shock finding algorithm identifies this front since it is nearer
    to the interface.

    \paragraph{Calculating Shock Speeds}
    The ``lab-frame'' shock front speeds are measured directly from computed radii 
    using a second-order-accurate finite-difference solution for the 
    first derivative accessed through the \texttt{numpy.gradient()} function.
    We refer to the lab-frame shock speeds as $\dot{R}_\mathrm{fwd/rev}$ 
    (forward/reverse shock).
    The calculated velocities ``jitter'' due to the discretization of the temporal
    and spatial domains.
    

    \paragraph{Calculating Ejecta and CSM Masses}
        
        The density is assumed constant in each cell, and is
        multiplied by the volume of the cell to get the mass of
        each cell in the domain. 
	    \PaulsCode\ gives the cell extent as an output and the
        cell volume is calculated as $V_i = 4\pi/3 (r_i^3 - (r_i-\Delta r_i)^3)$.
        To separate the mass into ``CSM'' and ``ejecta'', the cell
        masses are multiplied by the CSM fraction in the cell
        (ejecta mass calculated by subtracting CSM mass from total). 
        The cell masses can then be summed, e.g., only over the 
        cells of shocked gas to get the total mass of shocked material.

    \paragraph{Fitting to Simulation Data}
        Fits to simulation data described in the analysis were 
        carried out with the \texttt{scipy.optimize.curve\_fit()} routine.

\subsection{Evolution of forward shock radius}\label{ssec:shock_r}
%
%

\begin{figure}
            \centering
            \includegraphics[width=\linewidth]{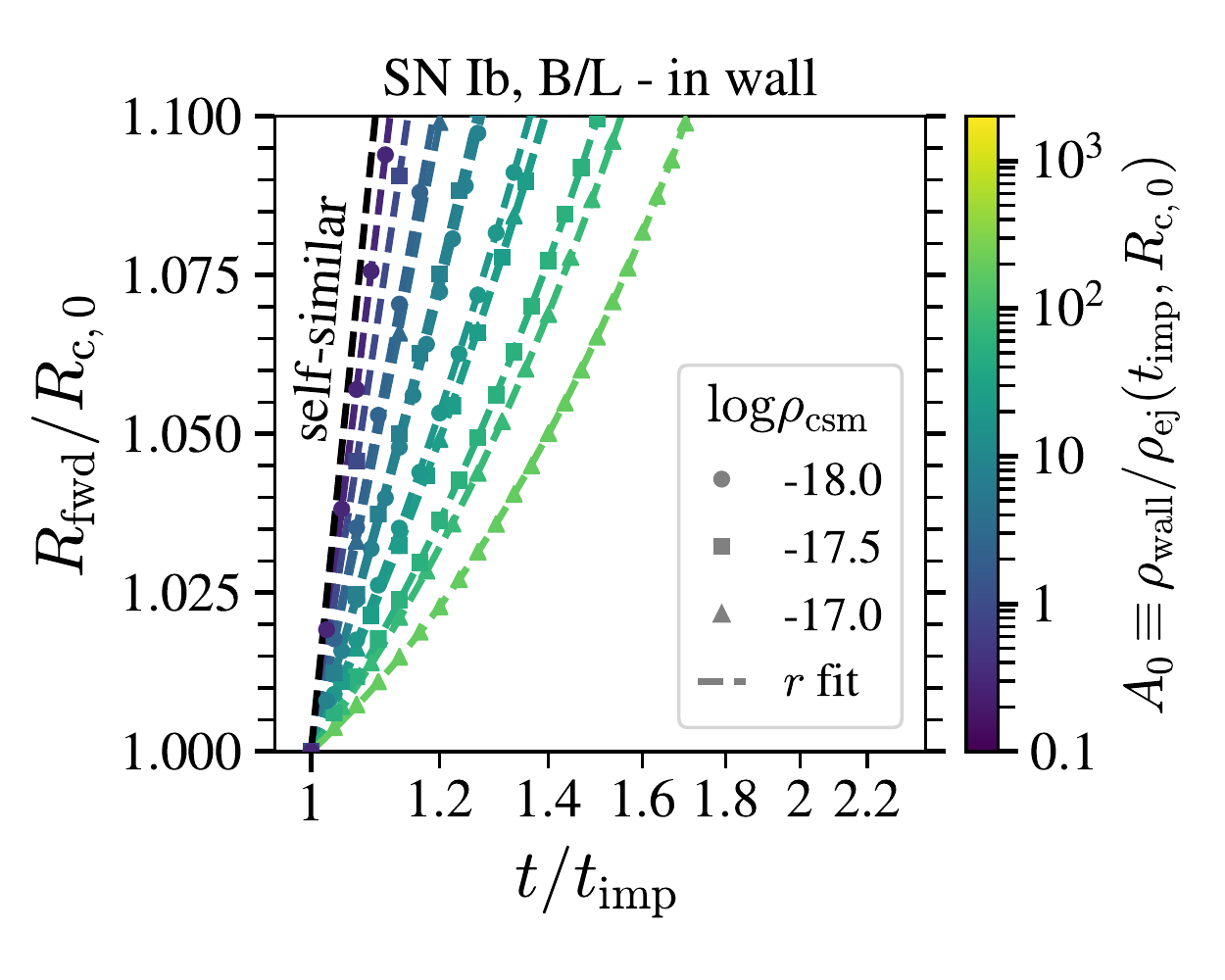}
            \caption{Evolution of forward shock radius while 
            within the wall for SN\,Ib models ($A_0=0.3-188.6$).
            Marker shapes denote wall density, 
            in units of $\densu$ (value in legend), 
            to show that $A_0$ is the governing parameter of the curve shape.
            Dashed lines in the $A_0$ color show the fit discussed in the text.
            The dashed black line shows self-similar evolution for reference.}
            \label{fig:r_sh}
        \end{figure}

Figure \ref{fig:r_sh} shows the evolution of forward shock radius 
while it is inside the wall for baseline SN\,Ib models. 
Models of different $\eta$ and $s$ are not shown, 
as the curves would be the same.
Color corresponds to the $A_0$ parameter.
Markers show the time of snapshots and their shape denotes the 
density of the wall. 
The black dashed line shows the self-similar solution, which
the lowest-$A_0$ models are approaching.
The remaining dashed lines show a fit to these data, described below.

The governing parameter for the evolution of shock radius is $A_0$. 
It is intuitive that a higher-density wall would lead to a slower shock.
But one must be careful, what matters is not the absolute density of the 
wall but rather its density {\it compared to} the ejecta density --- 
note from the marker shapes that walls of different absolute densities
can have the exact same $R_\mathrm{fwd}(t)/\Rsw$ 
if they have the same $A_0$. 

\begin{figure}
            \centering
            \includegraphics[width=0.45\linewidth]{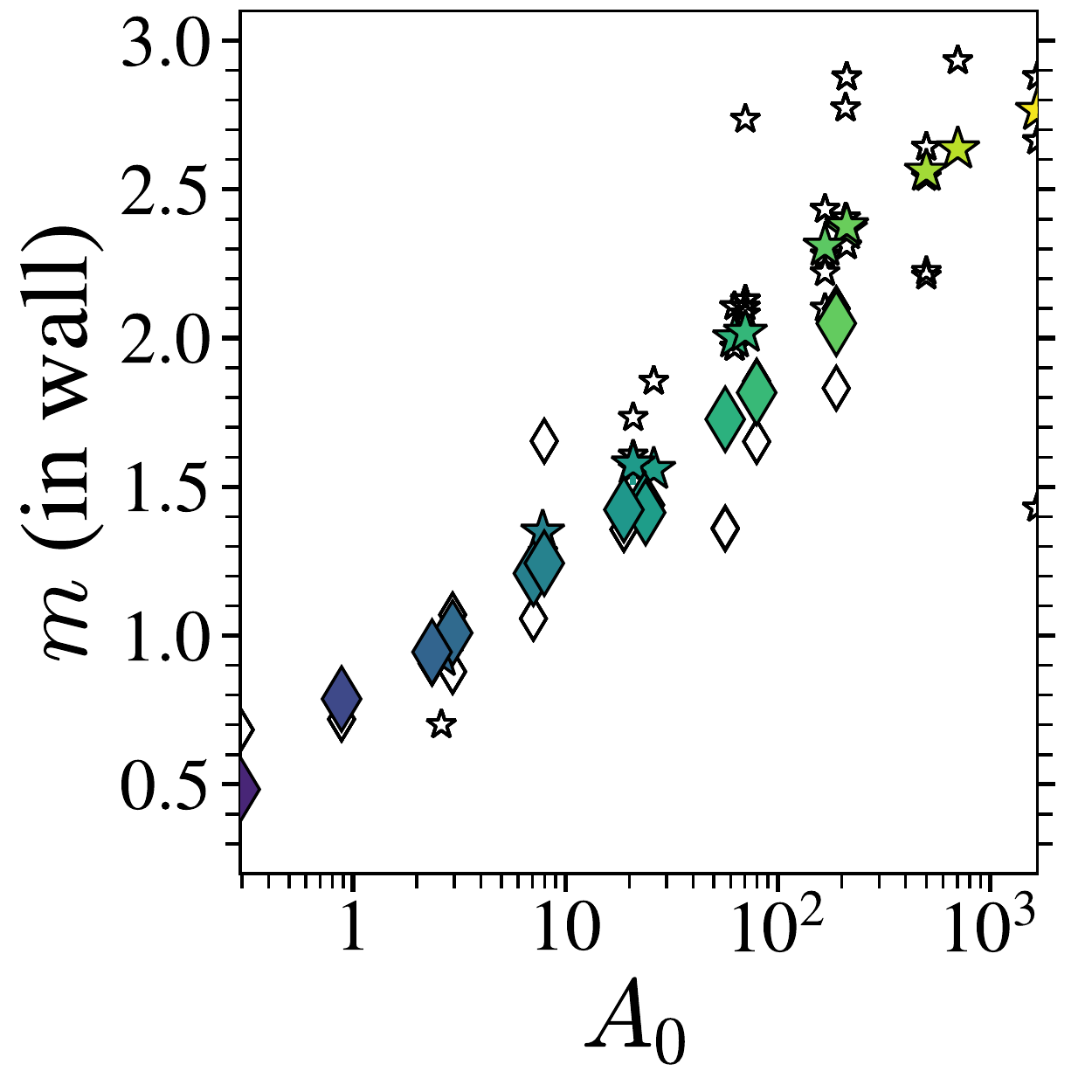}
            \includegraphics[width=0.45\linewidth]{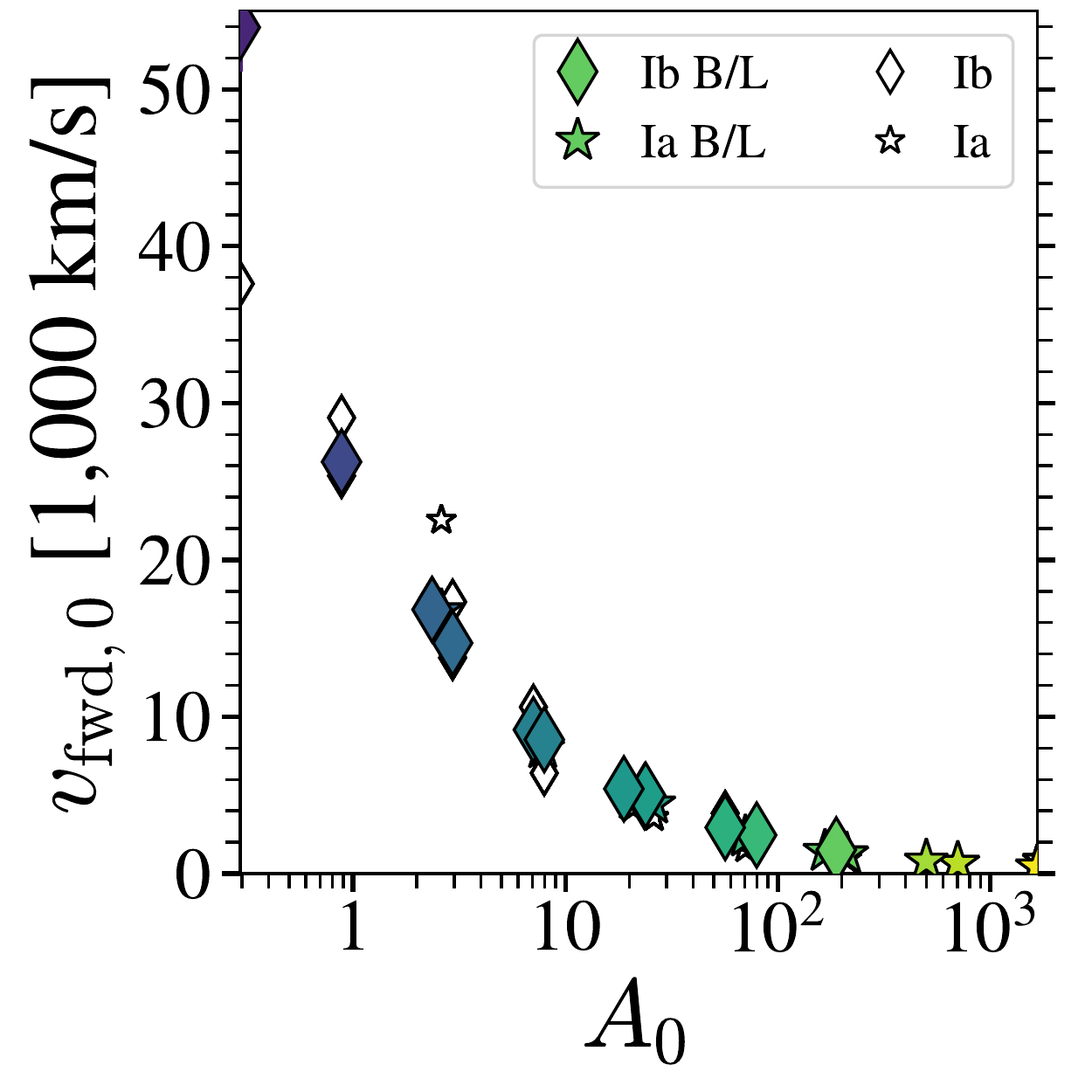}
            \caption{Fit parameters for $\Rfwd(t)$ as a function of $A_0$ 
            for the full model suite. 
            Outer CSM parameters are irrelevant for this phase of the evolution,
            so scatter reflects temporal and spatial resolution.}
            \label{fig:r_sh_plaw}
        \end{figure}

These curves can be described by an integrated power-law whose
parameters depend on $A_0$. 
In \S~\ref{ssec:shock_v} we demonstrate that the evolution of forward
shock velocity while in the wall is nearly a power-law
$v_\mathrm{fwd} = v_\mathrm{fwd,0} (t/\timp)^{m-1}$, 
with $m$ depending on $A_0$. 
Then $\Rfwd(t)$ should be described by its integral,
\begin{equation}
\frac{\Rfwd}{\Rsw} = \frac{v_\mathrm{fwd,0}}{\vejmax}\left(\frac{t}{\timp}\right)^m 
+ \left( 1-\frac{v_\mathrm{fwd,0}}{\vejmax} \right)~,
\label{eqn:Rfwdwall}
\end{equation}
where we have used the boundary condition $\Rfwd(t=\timp) = \Rsw = \vejmax \timp$.
Fitting this function to the data we can extract the best-fit
values  of $v_\mathrm{fwd,0}/\vejmax$ and $m$
(terms are collected in the fitting equation).
The best-fit values for each simulation in the suite are shown in 
Figure~\ref{fig:r_sh_plaw} as a function of $A_0$, limited to fit
results that gave a maximum deviation between the data and fit of 
$<0.1\%$, which we found is effectively a cut on spatial resolution. 
The left panel shows the power-law slope $m$ and the right panel shows the 
initial shock speed -- recall that $\vejmax$ is fixed to $30,000~\kms$ for our simulations
(\S~\ref{sec:models}). 
In these panels, diamonds and stars denote SN\,Ib and SN\,Ia models, respectively.
The large, filled markers are from the high-resolution baseline set, with fill color
corresponding to $A_0$ simply to further highlight them in the figure.
For these points we show the fit errors, the square root of the covariance matrix diagonal 
elements, though they are typically smaller than the marker size.
Unfilled, smaller markers represent models outside the baseline set. 
We include them to investigate the scatter caused by spatial resolution --
since this fit is only to the in-wall portion of the shock evolution, the variation
in $s$, $\eta$, and $F_R$ should not affect the fit results.
As $A_0$ decreases, $v_\mathrm{fwd,0}$ increases, and is 
higher than $\vejmax$ for $A_0 < 1$ (ejecta higher density than wall).
As can be seen from the way we expressed the $\Rfwd/\Rsw$ function, 
the result of higher $v_\mathrm{fwd,0}$ is that 
the second term disappears and the shock radius evolves as purely a 
power-law in time, just like in the self-similar case.
In fact, it is for this reason that we have chosen the variable $m$, 
since it is common in self-similar evolution applications to 
write $R\propto t^m$, and refer to $m$ as the ``deceleration parameter.''

In future, one may wish to use the $m$ and $v_\mathrm{fwd,0}$
values shown in Figure~\ref{fig:r_sh_plaw} and Equation~\ref{eqn:Rfwdwall}
-- or its derivative -- 
when interpreting observations (example in \S~\ref{ssec:14Crev}).
As a convenient alternative to reading $m$ and $v_\mathrm{fwd,0}$ values
off these plots, we offer the approximations
$m = 0.86 A_0^{0.177}$ and $v_\mathrm{fwd,0} = (25,600~\kms)\,A_0^{-0.56}$,
which we found by fitting first-order polynomials to the base-10 logarithm
of the quantities, fitting both SN types together since there is not more 
scatter between types than within a given type.

\begin{figure}
            \centering
            \includegraphics[width=\linewidth]{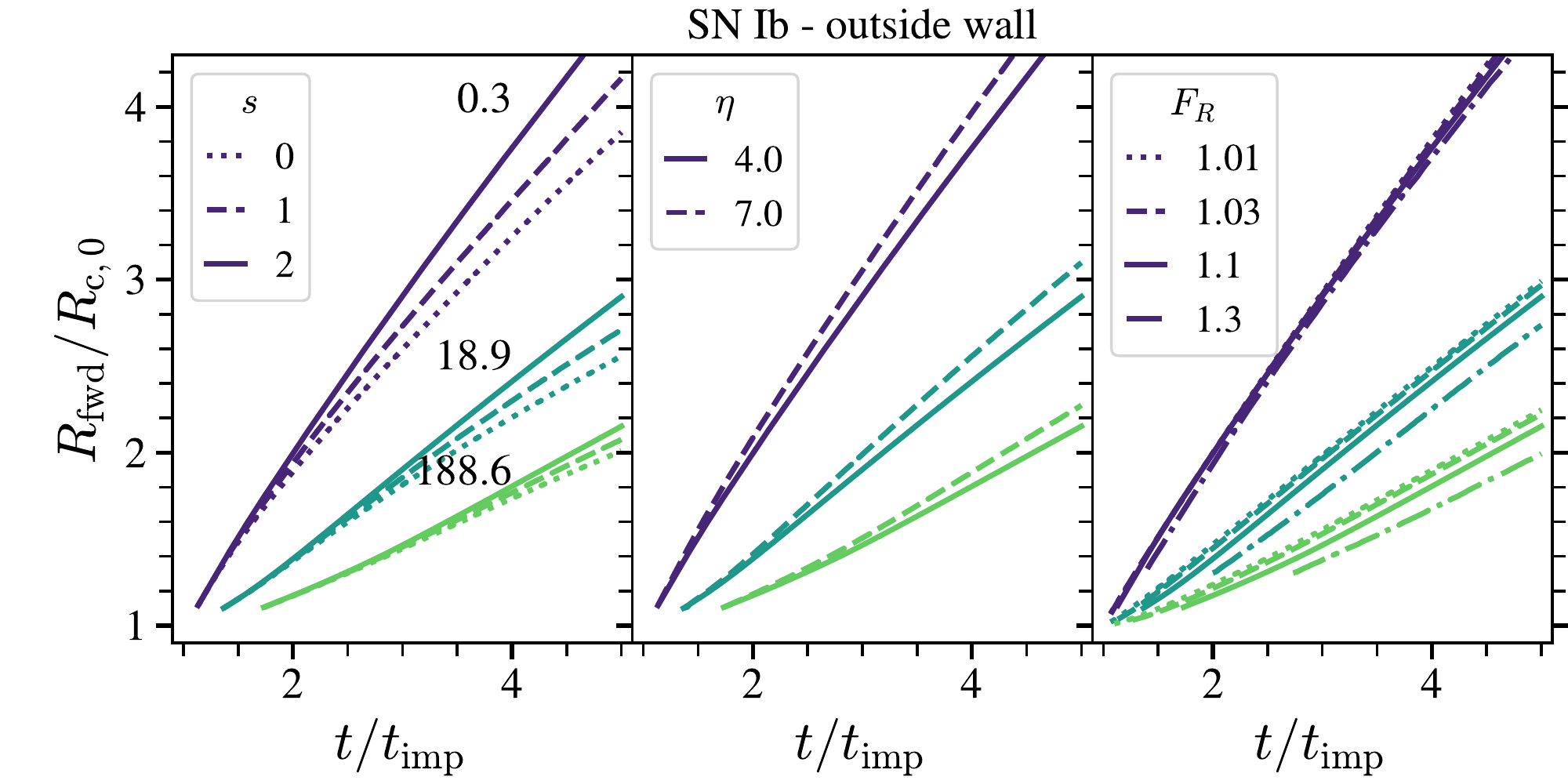}
            \caption{Evolution of forward shock radius while in the 
            after $\tx$, changing the three outer CSM parameters in turn.
           Solid lines are baseline values.
            }
            \label{fig:r_sh_svar}
        \end{figure}

Figure~\ref{fig:r_sh_svar} shows the evolution of the forward 
shock radius while the shock is in the outer CSM, varying 
the outer CSM properties $s$, $\eta$, and $F_R$ in turn
about baseline values $s=2$, $\eta=4$, and $F_R=1.1$.
For clarity, only the maximum, minimum, and median values of $A_0$
are shown in this plot with values annotated in the left panel.

\begin{figure}
            \centering
            \includegraphics[width=\linewidth]{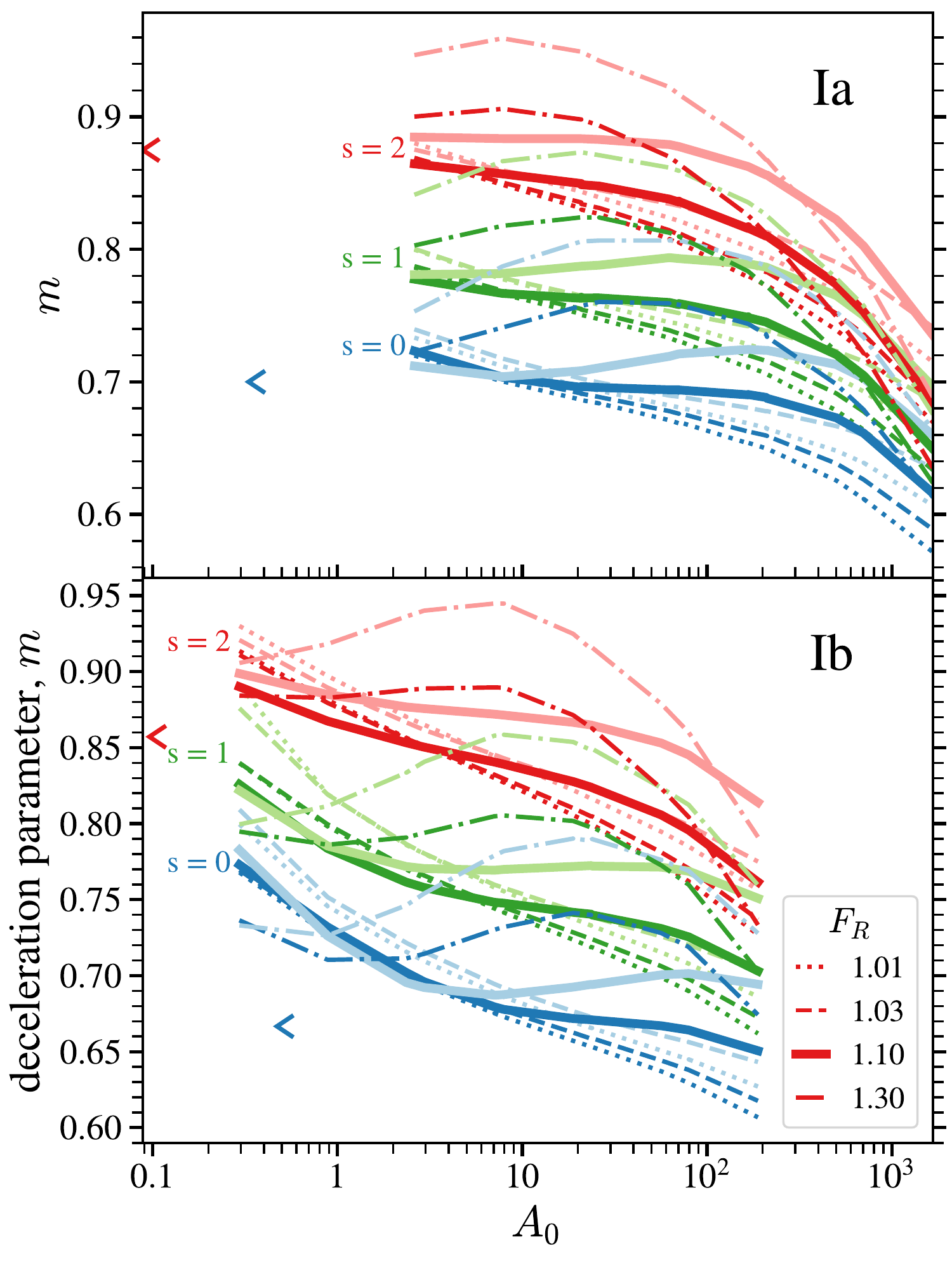}
            \caption{Power-law index $m$ of $R_\mathrm{fwd}\propto t^m$
            found by fitting the $3\leq t/\timp \leq5$ phase of 
            the radius evolution, as a function of $A_0$, for the entire suite 
            of SN\,Ib (bottom) and SN\,Ia (top) models.
            Line style shows $F_R$ (see legend), 
            color family indicates $s$ (see annotations),
            and saturation denotes $\eta$ (ligher shades are $\eta=7$).
            Arrows indicate the mini-shell value.
            }
            \label{fig:r_sh_svar_plaw}
        \end{figure}

By $t/\timp = 3$, models can be fit by 
$\Rfwd\propto t^m$, but $m$ depends on the CSM parameters.
To investigate this systematically and determine which parameters have
the maximum effect, we 
fit all models in the suite at times $t/\timp\geq3$.
The median difference between the power-law fit and the actual data
is $<1\%$ for all models.
The result is shown in Figure~\ref{fig:r_sh_svar_plaw}.
In this figure, color family shows $s$ (red is $s=2$; green, $s=1$; 
blue, $s=0$), saturation shows $\eta$ (light is $\eta=7$, dark is $\eta=4$), 
and line style shows $F_R$ (see legend).
The span of the plots is the same to allow for comparison between the
SN types.
The C82 values for the power-law slope are shown for $s=0,2$
as right arrows for reference.
For $s=0$ the arrows can be plotted at $A_0=A$, but for $s=2$, $A=0.096, 0.067$ 
(for SN\,Ib, Ia) are outside the span of the plot and are plotted at the left edge.

As $A_0$ increases, the 
parameters become degenerate. 
Yet the deceleration parameter is a useful probe of the 
CSM (or ejecta) density profile when $A_0$ is low -- in that case, different
values of $\eta$ and $F_R$ converge and $m$ values are separated by $s$.
However, by $A_0\sim10$ the separation is lost, and a 
single value of $m$ can be traced to a variety of CSM parameters, 
thus negating the interpretive importance of $m$ alone.

%
%
\subsection{Width of the Shock Region}\label{ssec:shock_dr}

\begin{figure}
            \centering
            \includegraphics[width=\linewidth]{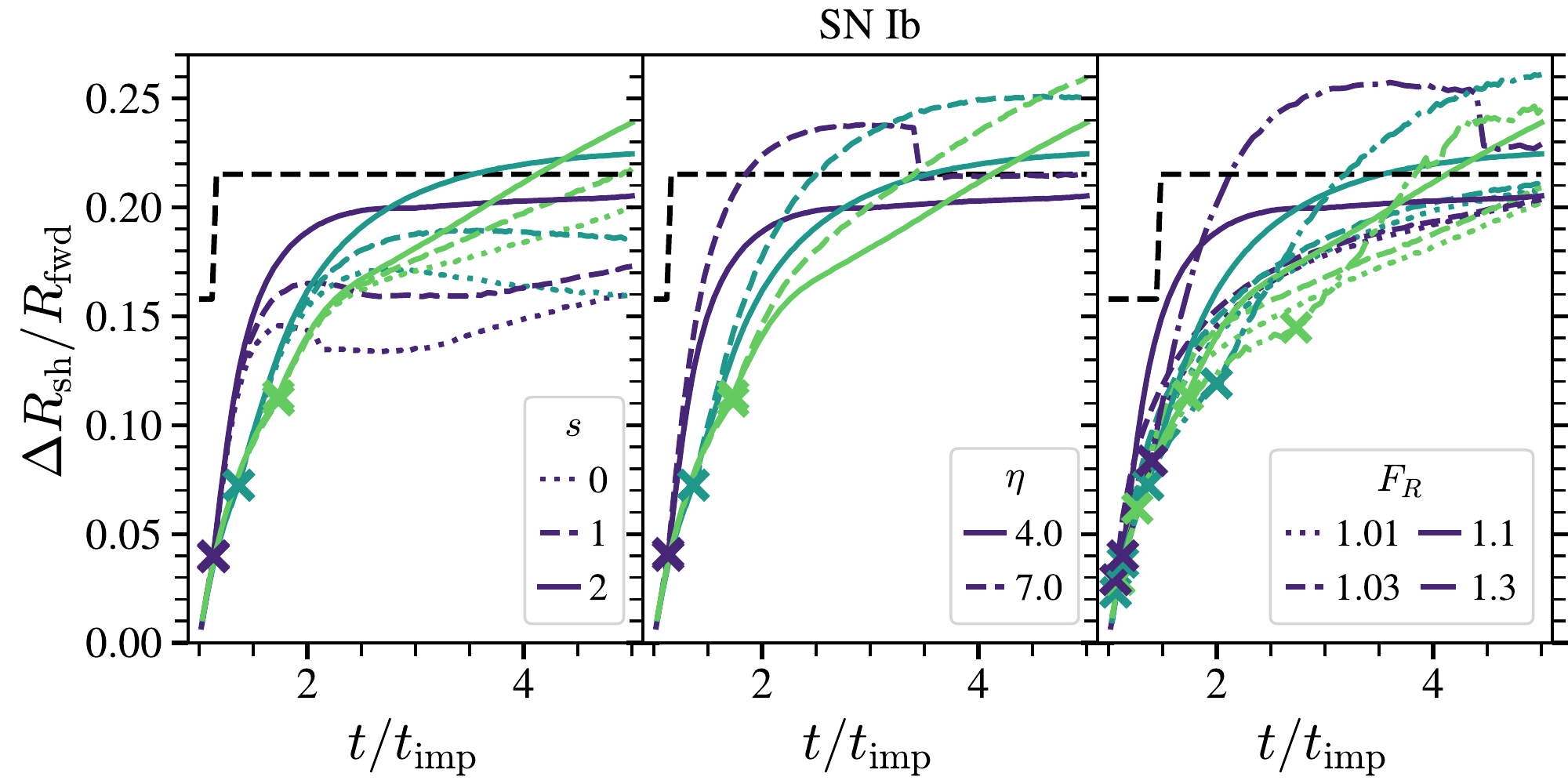}
            \caption{Fractional width of the shock region.
            Cross markers indicate $\tx$.
            Panels and colors are as in Figure~\ref{fig:r_sh_svar}.
            There are no clear trends with variations in CSM 
            configuration, the shock width quickly grows to $\sim0.2\Rfwd$
            for all models.
            Self-similar value shown as black dashed line.
            }
            \label{fig:thickness}
        \end{figure}

One key feature of self-similar evolution is that the shock front locations
are a constant multiple of the contact discontinuity radius
and therefore the fractional width of the shock 
\begin{equation}
\frac{\Delta R}{\Rfwd} = 1 - \frac{\Rrev}{\Rfwd}
\end{equation}
is also a constant.
Figure~\ref{fig:thickness} shows that in our models, 
however, this is not the case. 
This figure is structured the same way as  Figure~\ref{fig:r_sh_svar},
with color showing $A_0$, line style indicating different values 
of a CSM parameter, and different CSM parameters are changed in 
each panel.
The full time span ($1 \leq t/\timp \leq 5$) is shown, 
and crosses show $\tx$.
Black dashed lines show the self-similar values.

We find that the shock width generally grows quickly at first,
but settles to $\sim 20\%$ by $t/\timp\sim2$.
There is no obvious dependence of the shock width on $A_0$, 
$s$, $\eta$, or $F_R$. 
Interpretation of observations prior to the asymptotic phase
should account for the fact that the shock region is likely
much thinner even than the mini-shell may predict, 
but at later times the typical assumption of $\sim10\%$ 
is a decent approximation.

For some low-$A_0$ models, the thickness seems to drop suddenly
at $t/\timp\sim3$.
This is because of the shock front finder identifying the rarefaction
shock front instead of the reverse shock front,
as described in \S~\ref{ssec:an_meth}.

\subsection{Shock Speeds}\label{ssec:shock_v}

\begin{figure}
            \centering
            \includegraphics[width=.75\linewidth]{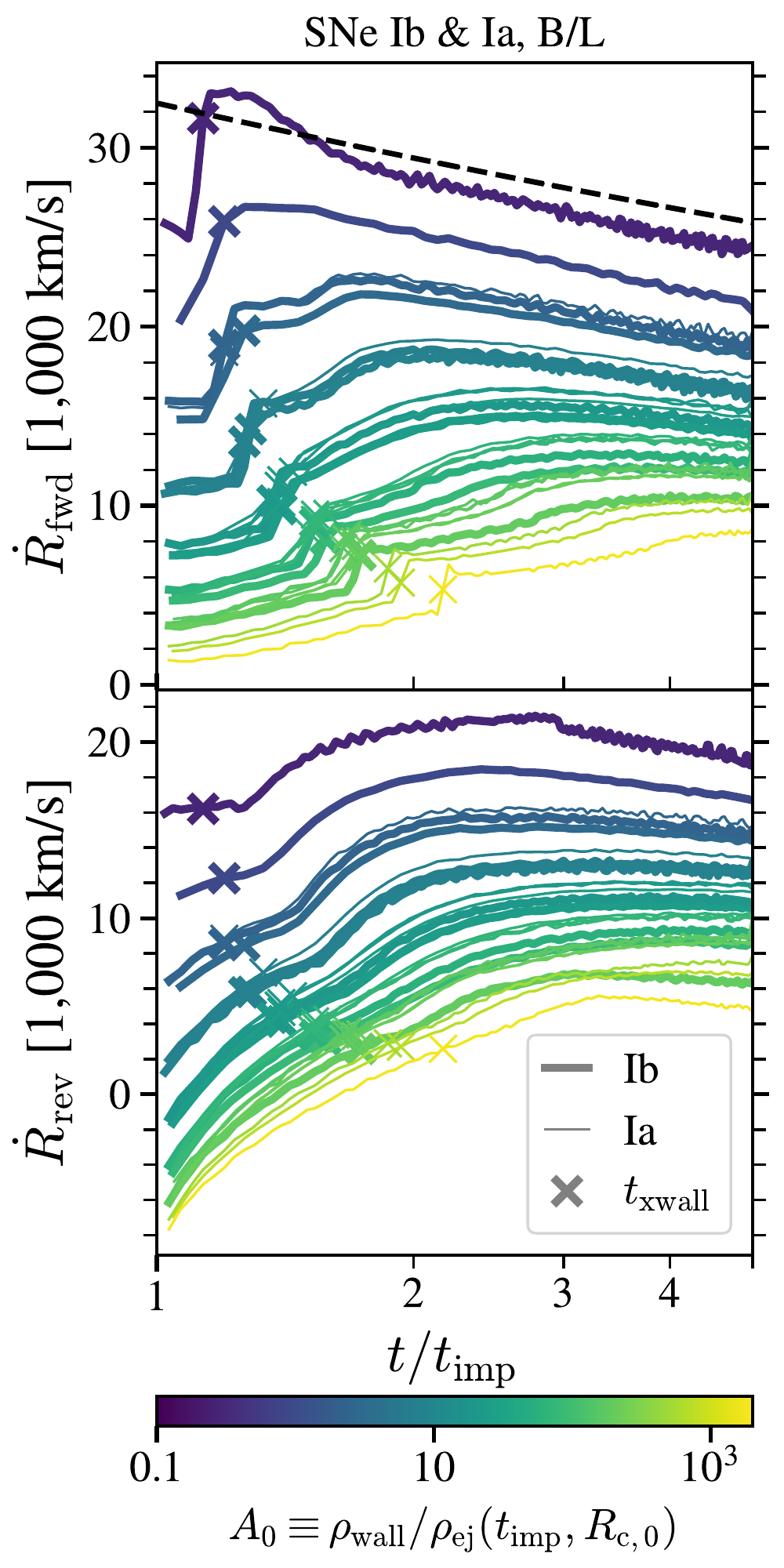}
            \caption{Forward (top) and reverse (bottom) shock speeds 
            for baseline CSM models.
            Wall crossing time is indicated by an ``x.''
            The black dashed line in the top panel shows the self-similar
            value for the lowest-$A_0$ simulation.
            }
            \label{fig:v_sh}
        \end{figure}

Figure \ref{fig:v_sh} shows the evolution of the forward (top) 
and reverse (bottom) shock speeds
throughout the duration of the simulation for baseline SN\,Ia \& Ib 
models (thin and thick lines, respectively).
The time that the shock crosses into the outer CSM, $\tx$, 
is marked by an ``x''.
Curves are color-coded by $A_0$, with highest $A_0$ having the lowest 
shock speeds. 
The black dashed line shows the self-similar solution for the CSM
of the lowest-$A_0$ model.

In contrast to self-similar evolution, most models actually have
an accelerating forward shock in the wall.
In the self-similar regime, shocks do not accelerate unless $s>3$.
The forward shock speed while the
shock is in the wall is fit very well by a power law
\begin{equation}
v_\mathrm{fwd} = v_\mathrm{fwd,0} \left( \frac{t}{\timp} \right)^{m-1} ~.
\end{equation}
The $m$ derived from fitting this function to the in-wall portion of
$v_\mathrm{fwd}$ yields the same results as presented in 
Figure~\ref{fig:r_sh_svar_plaw}
from fitting its integral to the radius evolution  
-- in \S~\ref{ssec:shock_r} we offer power-law fits for 
$m(A_0)$ and $v_\mathrm{fwd,0}(A_0)$, for convenience.
The shock is accelerating ($m>1$) for $A_0 \gtrsim 10$.

After $\tx$, the forward shock is in the outer CSM, which
for the baseline set has a wind-like density profile.
The self-similar line describes the velocity evolution in the outer
CSM very well for low walls (low $A_0$) but does not suit well for 
high walls until later times.
As a rule of thumb, 
we find that self-similar evolution may be used at 
times later than the time at which $\rhoej(\Rsw)=\rhowall$.

\subsection{Deceleration of Ejecta}\label{ssec:ej_v}

\begin{figure}
            \centering
            \includegraphics[width=\linewidth]{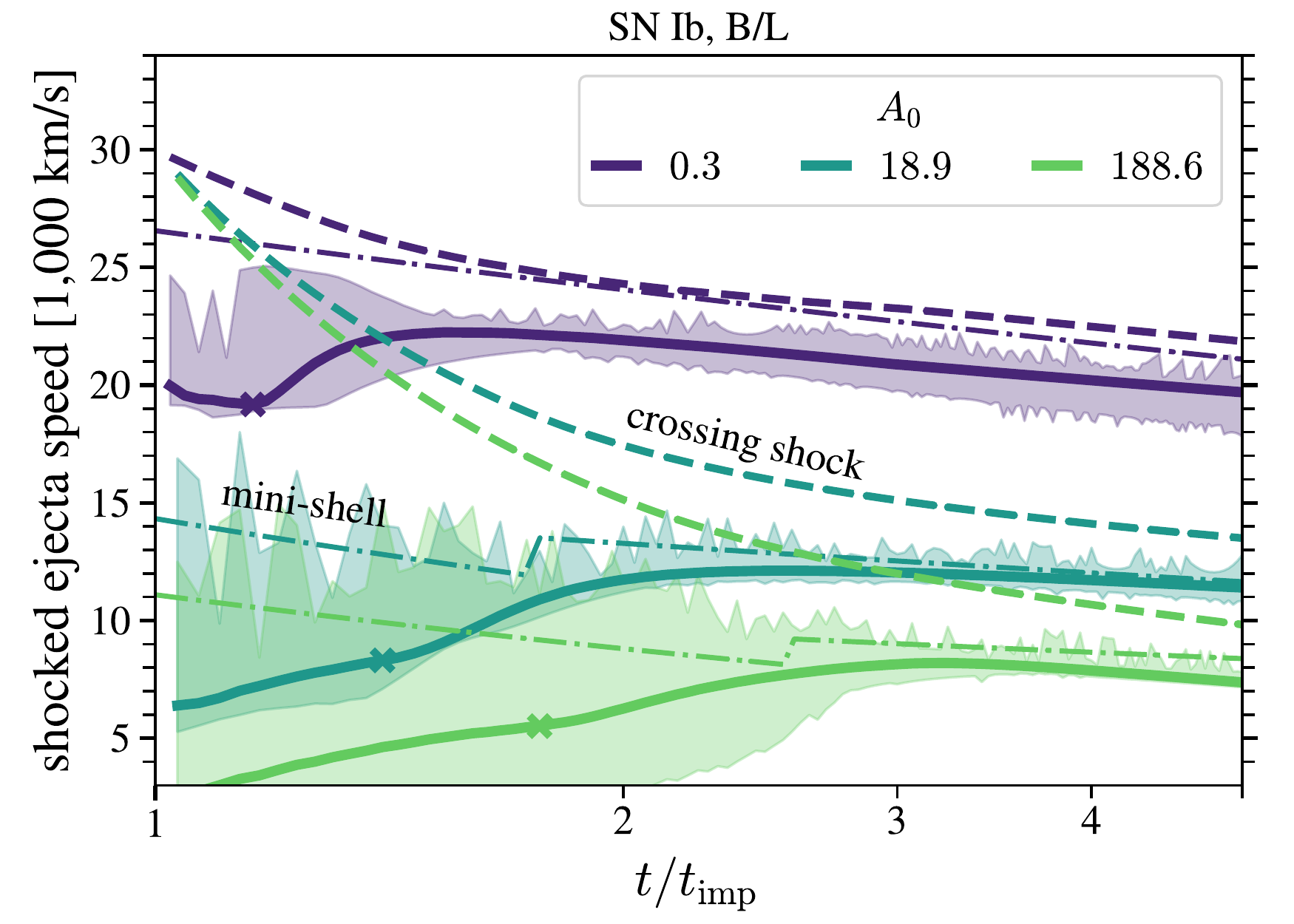}
            \caption{Shocked ejecta speed over time for three  $A_0$
            values -- baseline SN\,Ib.
            Solid lines are volume-average and envelopes show min./max.
            Dashed curves show the speed of the ejecta crossing the shock
            front.
            Dash-dotted curves show the mini-shell value.
            Time $\tx$ is marked by ``x''. 
            Higher-$A_0$ walls decelerate the ejecta significantly more
            than low-$A_0$ walls and more than self-similar.
            }
            \label{fig:vej}
        \end{figure}

Figure~\ref{fig:vej} shows the evolution of shocked ejecta speed
($\uej$)
over time for three baseline models 
representing the span of $A_0$ -- minimum, median, and maximum values.
In case the colors denoting $A_0$ are not distinct to the reader, 
we note the lowest $A_0$ value is always the highest-speed curve
(for any line style). 
The shaded band shows the minimum and maximum values of $\uej$
across cells of shocked ejecta.
Noise in these curves is an effect of the steep velocity gradient 
near the reverse shock front and the resolution of the simulations. 
To minimize numerical noise we focus on the volume-average shocked 
ejecta speed, $\langle \uej \rangle_{V}$, shown as a solid curve.

The stopping power of the wall can be illustrated by a comparison 
of the shocked ejecta speed to the speed of the ejecta crossing into 
the shock region ($R_r/t$, dashed lines).
It is obvious that a higher wall (higher-$A_0$ model)
is able to slow the ejecta much more than a lower wall. 
Given that the CSM speed is $100~\kms$, the highest wall has effectively stopped
the ejecta.
After the forward shock has traversed the wall, however, the 
wall rarefies and moves outward and so the ejecta are able to
be accelerated by the fast material still pushing from behind.
The shocked ejecta reach some maximum speed $u_\mathrm{ej,max}$
and then begin to decelerate.

In fact, once in the decelerating regime the shocked ejecta speed is 
similar to the mini-shell prediction for evolution in the outer CSM
density profile (dash-dotted lines),
which can be calculated from Equation 21 of C82.
Here, we will deviate from the original notation by 
using subscripts ``f'' in place of ``1'' (forward shock) and 
``r'' in place of ``2''(reverse shock). 
We assume $R_r/t$ for the pre-shock velocity of the ejecta.
The variable $u_{r/f}$ will be used to refer to post-shock gas speed
at the reverse/forward shock front.
Using the constancy of $u_r/u_f$, $R_c$ given by Equation 3 of C82,
and $R_r/t = (R_r/R_c)(R_c/t)$, $u_r$ is found to be:
\begin{eqnarray}
u_r &=& \frac{3}{4}\left( \frac{n-3}{n-s} \right)
\left(\frac{u_r}{u_f}\right)
\left(\frac{R_f}{R_c}\right)
\left(\frac{Ag^n}{q} \right) ^{\frac{1}{n-s}} t^{(s-3)/(n-s)}
\end{eqnarray}
The parameter $A$ is fixed for given $n$ and $s$.
The variables $g^n$ and $q$ are the normalizations of the density profiles, 
as in \S\,\ref{sec:models}.
The cautious reader may note from C82 that 
gas velocity decreases between the reverse and forward shocks, 
so $\langle \uej \rangle_{V} < u_r$.
However, due to the thinness of the reverse shock region
($[R_c-R_r]/[R_f-R_r]=0.063$), $\uej$ is effectively constant
in the mini-shell model.
For $s<3$ the ejecta will be constantly decelerated by
the CSM and therefore the maximum speed is achieved at $t/\timp=1$, 
unlike in our models whose maximum speed is later even than $\tx$.
As in \S~\ref{ssec:shock_v}, peak speed occurs when the 
reverse shock reaches ejecta of approximately the wall density.

Emission line widths for interacting SNe are often used as a 
proxy for the bulk velocity of shocked gas 
\citep[e.g., for delayed interaction,][]{Dilday+12,Milisav+15}.
Figure~\ref{fig:vej} shows that for delayed-interaction SNe, 
one may observe significant deceleration of ejecta by 
a modest mass of CSM that is nevertheless much more dense than
the ejecta impacting it. 
This principle is independent of the exact CSM configuration
we are presenting in this work: low line velocities 
do not necessarily indicate a CSM mass comparable to or exceeding 
the total ejecta mass, in delayed-interaction events.
In the context of our model suite, the observed line widths of 
a few $1,000~\kms$ or lower would suggest that delayed-interaction 
SNe tend to occur with $A_0\gtrsim100$.

\subsection{Mass of Shocked Material}\label{ssec:shock_m}

\begin{figure}
            \centering
            \includegraphics[width=\linewidth]{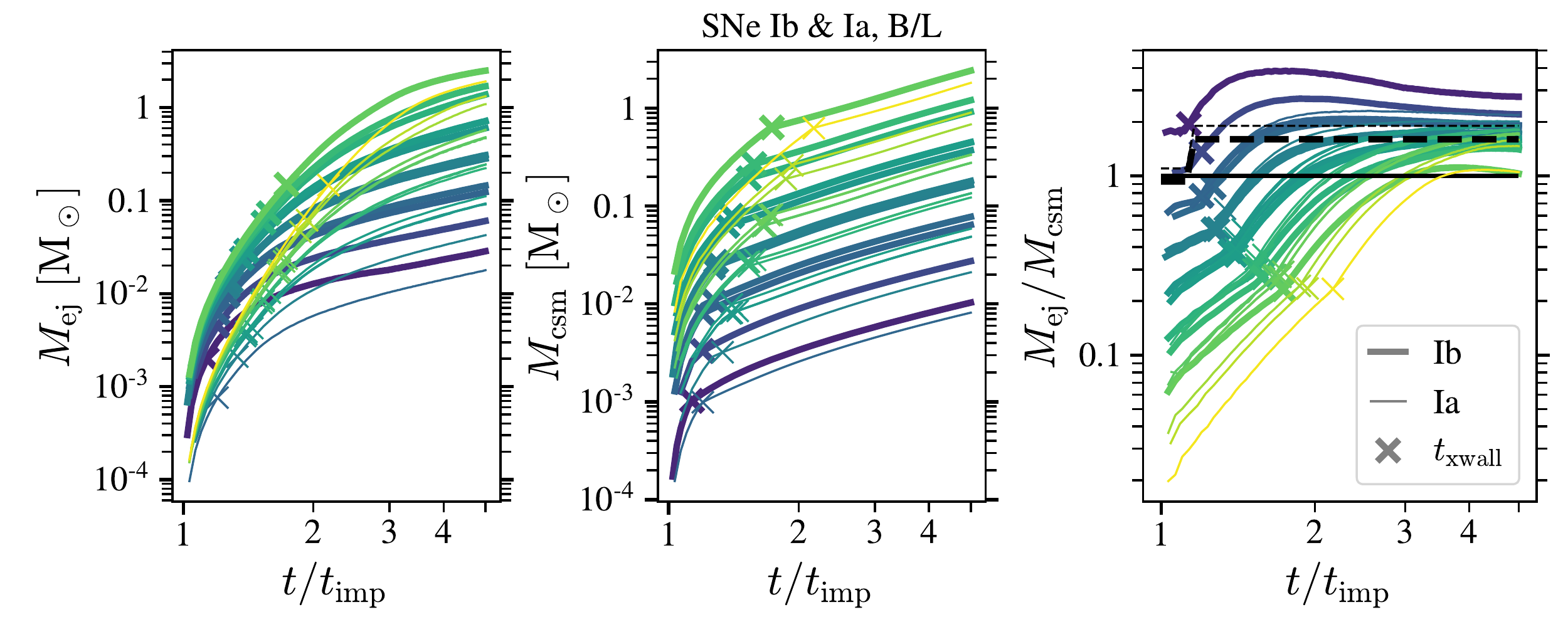}
            \caption{
            The masses of shocked ejecta (left) and CSM (center),
            and the ratio of shocked eject mass to shocked CSM mass (right).
            In the right panel, a solid black line shows unity and
            dashed black line shows the self-similar value.
            In all panels, ``x'' marks $\tx$.
            }
            \label{fig:M_sh}
        \end{figure}

Figure~\ref{fig:M_sh} shows how the masses
of shocked ejecta (left) and CSM (middle) evolve with 
time, for baseline CSM models.
Color indicates $A_0$, the initial density ratio between
the CSM wall and the ejecta.
Higher-mass curves have higher $A_0$ values. 
Time is normalized to time of impact.        
The time the forward shock crosses the wall, $\tx$,
is marked by an ``x.''
The right panel shows the ratio of shocked ejecta mass
to shocked CSM mass (a proxy for the average composition of 
shocked material) with unity marked by a solid black line.
The self-similar value is shown as a black dashed line.

The left and middle panels show that, for a given SN type,
the mass swept up is the same for the same $A_0$ values even 
though they represent different wall densities. 
This is because a lower-density
wall farther away has a larger volume.
Lower-wall models (lower $A_0$) have lower shocked masses.

The mass of the wall is
\begin{equation}
    M_\wall = (2.1\times10^{-3}~\Msun)~\rho_\mathrm{wall,-18} R_\mathrm{c,0,16}^3 (F_R^3 - 1)
    \label{eqn:Mwall}
\end{equation}
where $\rho_\mathrm{wall,-18} = \rhowall/(10^{-18}~\densu)$, 
$R_\mathrm{c,0,16} = \Rsw/(10^{16}~\cm$, and 
$F_R$ can be either assumed or 
found by using $t=\tx$ in Equation~\ref{eqn:Rfwdwall}, 
recalling that $m$ and $v_\mathrm{fwd,0}$ depend on $A_0$ (\S~\ref{ssec:shock_r}), 
which is set by $\rhowall$ and $\Rsw$ (Equation~\ref{eqn:A0}).

Looking at how much of the shocked gas is ejecta (right panel),
we see that the highest walls have the lowest ratio of 
ejecta to CSM mass. 
All models converge to an ejecta-to-CSM mass ratio of $\sim$few
by late times.
At early times, for high walls it is not a good assumption that
approximately as much ejecta has been shocked as CSM -- it can be up 
to $t/\timp=3$ before that is the case.

\subsection{Effect of Mixing}\label{ssec:mixing}
    
    Unlike previous efforts to model interaction with CSM configurations
    similar to those we model, our simulations use a one-dimensional code
    that captures the mixing of ejecta and CSM in a way that 
    approximates the behavior of three-dimensional shocks.
    The simulations are of course limited compared to nature in that they 
    cannot reproduce the two-phase medium of hot and cold gas that the
    Rayleigh-Taylor instability is supposed to create and can impact
    observational signatures.
    
\begin{figure}
            \centering
            \includegraphics[width=\linewidth]{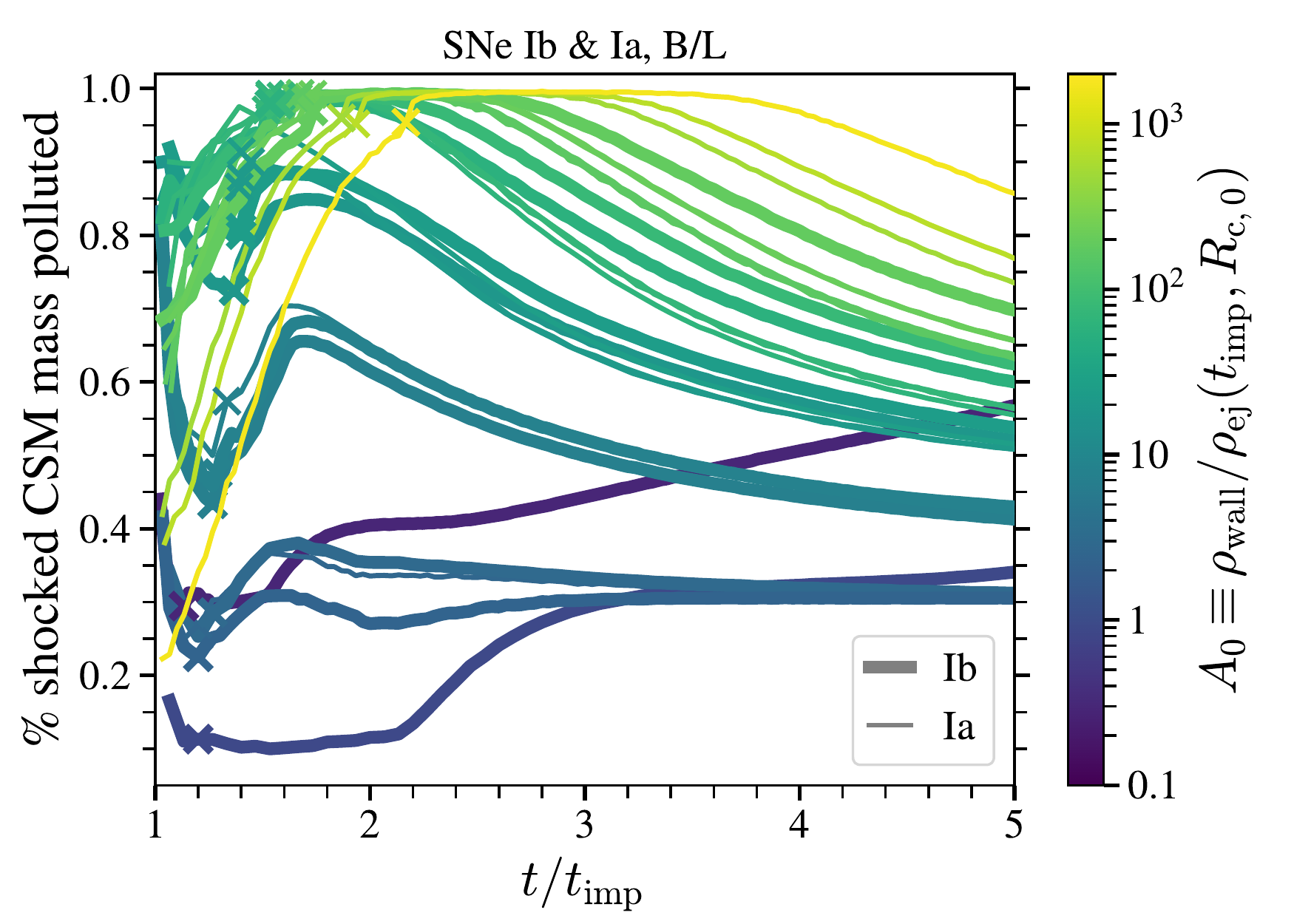}
            \caption{Fraction of shocked CSM mass that is a blend of 
            ejecta and CSM. 
            Crosses mark $\tx$.
            }
            \label{fig:mixed_frac}
        \end{figure}

    Interesting for observations is how polluted the CSM becomes from 
    ejecta, which may have higher metallicity.
    For this analysis we consider ``pure ejecta'' to be gas with
    a CSM mass fraction $X < 0.01$ 
    and, likewise, ``pure CSM'' has $X > 0.99$.
    Thus, ``polluted CSM'' is gas with $0.01 \leq X \leq 0.99$. 
    Figure~\ref{fig:mixed_frac} shows, for baseline CSM models,
    the percentage of shocked CSM mass 
    that is polluted, $100\times M(0.01\leq X \leq0.99)/M(X\geq0.01)$.
    Thick lines show SN\,Ib models and thin lines show SN\,Ia models.
    Color is $A_0$ and ``x'' shows the time the forward shock
    crosses the wall, as in other figures.

    We find interesting diversity in the evolution of CSM pollution.    
    Regardless of $A_0$, as the shock evolves in the outer CSM it
    will tend toward self-similar evolution with a healthy 
    Rayleigh-Taylor instability and at least some mixing;
    but as with other hydrodynamical properties, the amount of time
    it takes for the self-similar solution to apply depends on
    $A_0$.
    For the highest walls, mixing increases while the shock is
    in the wall, even causing all of the wall mass to be polluted.
    Mixing stops when the shock crosses into the outer CSM, but 
    the instability slowly gains strength again.
    For the lowest walls, mixing is unimportant in the wall phase but
    the instability grows as the shock traverses the outer medium.
    We see that $\sim60\%$ of shocked CSM is polluted material
    by the end of the simulation for the lowest wall, and that fraction
    is growing.
    In the range of intermediate wall heights, the behavior is somewhat
    complicated.
    Mixing seems to make a start when the shock initially crosses
    into the outer CSM, but is dampened, possibly by the rarefaction
    wave. 
    The percent of polluted CSM grows faster in the outer CSM 
    for lower walls.
    In summary, due to the different growth and damping timescales
    of mixing for different $A_0$, we find that
    high walls have mostly polluted CSM except at the earliest times, 
    low walls have low CSM pollution until late times, 
    and intermediate walls have moderate CSM pollution at all times.

    The only general rule we offer is that at least 10\% of shocked
    CSM seems to be polluted by ejecta material at the $>1\%$ level,
    across all $A_0$ and $t/\timp$.
    Thus when considering the composition of the shocked CSM 
    for interpreting observations of interactions, one needs
    to account for $A_0$ and the relative time of the observation,
    even for rough estimates.
    For any detailed analysis, of course, one would want to use
    the radial profile of the CSM fraction over time instead of
    the summary quantity we present here.

\section{Discussion}\label{sec:disc}

    \subsection{Application to Observations of SN\,2014C}\label{ssec:14C}
    Here we will apply our models to the observations of 
    SN\,2014C to demonstrate the application of our hydrodynamic 
    results for determining CSM properties.
    
    The properties of the SN ejecta are constrained 
    from the early light curve in \citet[][hereafter, M17]{Margutti+17}
    and our SN\,Ib model is set up to be consistent with 
    SN\,2014C.
    The key observations are the impact time from radio
    and optical data, 
    an estimate of the wall density from x-ray data,
    and measurements of the forward shock radius,
    deceleration parameter, and shock speed 
    from very long baseline radio interferometry (VLBI). 
    Putting these pieces together we constrain
    $A_0$ and $s$.

    The onset of interaction -- i.e., initial impact with outermost
    ejecta -- is not yet precisely constrained for SN\,2014C.
    Interaction certainly began by $120~\days$ post-explosion,
    based on the emergence of a $\sim1,000~\kms$ $\Hal$ emission
    line component in the optical spectra \citep{Milisav+15}, 
    and likely at about $100~\days$
    based on the flattening 15.7~GHz radio light curve
    \citep[][hereafter, A17]{Anderson+17}.
    For this discussion we also include $\timp=190~\days$
    (the assumption of both A17 and M17 based on the onset of 
    radio rise),
    though it is unclear to us how to reconcile this 
    $\timp$ with the strong $\Hal$ emission already at $120~\days$.
    This brings the final list of possible impact times to
    $\timp = 100, 120, 190~\days$. 
    In our models 
    $\Rsw = (2.56\times10^{16}~\cm)\timp/(100~\days)$, 
    so for the three $\timp$ we are considering,
    $\Rsw = (2.56, 3.07, 4.86)\times10^{16}~\cm$, which are all
    consistent with the x-ray non-detection limit of M17.

    We first attempt to estimate 
    a range of reasonable $A_0$ values for SN\,2014C
    via $\rhowall$ and $\timp$ (Equation~\ref{eqn:A0}).
    Using the x-ray emission measure at 500 days
    to derive the number of emitting particles, and the 
    volume of gas derived assuming self-similar evolution 
    of the shock, M17 derive 
    $\rhowall \sim 10^{-17}~\densu$.
    We note this value depends on the assumed composition,
    and while M17 assumed solar abundances, our modeling shows
    a significant degree of pollution from the ejecta 
    (Figure~\ref{fig:mixed_frac}).
    A hypothesis we will explore in the sequel paper
    on radiation signatures is that the M17 density is over-estimated,
    and here we will consider
    $\rhowall = 10^{-18}~\densu$ also possible.
    In our SN\,Ib suite,
    the model with $\rhowall = 10^{-17}~\densu$ and $\timp=120~\days$
    is the highest-$A_0$ model with $A_0 = 188.6$. 
    Scaling off of this using Equation~\ref{eqn:A0},
    for $\rhowall=10^{-17}~\densu$ the impact times give
    values of
    $A_0 = 110, 190, 750$ (rounding to the nearest ten).
    The $\rhowall=10^{-18}~\densu$ assumption has $A_0$
    a factor of ten lower than these values (11, 19, 75).
    Note that these values are all significantly above 
    $A_0\sim1$, indicating the self-similar solution 
    does not apply.
    One may notice that the inner radius assuming $\timp=190~\days$ 
    is nearer than derived in M17 
    ([4.9 vs.~5.5] $\times10^{16}~\cm$)
    because we define ``beginning of interaction'' differently --
    they assume the forward shock traversing the rarefied inner 
    cavity can be transmitted through the wall, whereas we assume
    it is negligible and wait for the ejecta to reach the wall -- 
    and if we used their  $\Rsw$ rather than their $\timp$, we would
    derive $\timp=215~\days$ and $A_0 = 1080$.

    For the nearest SNe, VLBI can be used to
    directly image the expanding shock fronts and measure 
    hydrodynamical quantities like $\Rfwd$, 
    $\Delta R_\mathrm{sh}$, and $v_\mathrm{fwd}$. 
    \citet[][hereafter, B18]{Bietenholz+18} present interferometry of SN 2014C
    $\sim400-1000$ days after explosion. 
    Given our range of $\timp$ values, the radio 
    interferometry spans $4 \leq t/\timp \leq 10$ (minimum $\timp$)
    or $2 \leq t/\timp \leq 5$ (maximum $\timp$).
    At these phases of evolution, nearly all of our models 
    have the forward shock already traversing the outer CSM,
    so the relevant figures for radial information are 
    Figures~\ref{fig:r_sh_svar}~\&~\ref{fig:r_sh_svar_plaw}.
    Measurements of $v_\mathrm{fwd}$ can be compared to the 
    baseline model set with Figure~\ref{fig:v_sh}.

    The first quantity of interest from the VLBI 
    measurement is the forward shock radius itself.
    B18 reports that at 384 days ($3.32\times10^7~\mathrm{s}$), 
    $\Rfwd = (6.4\pm0.3)\times10^{16}~\cm$, 
    corresponding to 
    $t_\mathrm{VLBI}/\timp = 3.84, 3.2, 2.02$ 
    and 
    $R_\mathrm{VLBI}/\Rsw = 2.5, 2.1, 1.3$.
    Looking at Figure~\ref{fig:r_sh_svar}, these values
    all lie around the $A_0=18.9$ model lines, independent of 
    $s$, $\eta$, or $F_R$.

    Next, we look at the measurement of the deceleration
    parameter.
    In \S~\ref{ssec:shock_r} we mentioned that the
    deceleration parameter, $m$, is used to describe
    the radial evolution as $\Rfwd \propto t^m$
    and showed that $\Rfwd(t/\timp > 3)$ 
    -- when the shock is in the outer CSM --
    can be fit precisely with a power-law, but $m$ depends on
    $s$, $\eta$, $F_R$, and $A_0$ (Figure~\ref{fig:r_sh_svar_plaw})
    and for even moderately high values of $A_0$, interpretation of 
    $m$ may be muddled.
    Fitting to the VLBI data, B18 find a 
    best-fit $m=0.79\pm0.04$ for SN\,2014C. 
    Despite the confusion in the high-$A_0$ region of Figure~\ref{fig:r_sh_svar_plaw},
    $s=0$ is disfavored --
    only $F_R=1.3$, $\eta=7$ models (with $A_0\lesssim 100$) 
    have $m>0.75$.
    Furthermore, assuming the sweeping up of the outer CSM
    occurred adiabatically and favoring $\eta=4$ models, 
    then $s=0$ is ruled out,
    only the $s=1$, $F_R=1.3$ models with $A_0 < 100$
    are consistent with the data, 
    but all $F_R$ values are within the margin of error
    for $s=2$.
    The wind-like outer medium favored by our model
    is in line with the analysis of \citet{Tinyanont+19}, who 
    found $s=2$ using the model of \citet{Moriya+13}, 
    which is self-similar but applies at these late times.
    Note that from the mini-shell solution 
    $m=(n-3)/(n-s)$ with $n=9$ and $m=0.79$, one would derive 
    $s=1.4$.
    Finally, for $\eta=4$, the measured range of
    $m$ suggests $A_0 < 300$ independent of $F_R$ and $s$,
    assuming the curves continue to decline.
    
    Finally, we can compare our models to the
    shock velocity measured by B18.
    Our models only capture their first 
    data point $v_\mathrm{fwd} = 14,500\pm3,400~\kms$ 
    at $t\sim514~\days$, which corresponds to
    $t/\timp \sim 5.1, 4.3, 2.7$.
    In our baseline suite ($s=2$, $\eta=4$, $F_R=1.1$, consistent 
    with the constraints on SN\,2014C from $m$), 
    models with $A_0\sim3-200$ match the 
    measured velocity given the measurement errors. 
    The B18 measurements show a nearly constant velocity, 
    which is most consistent with 
    $A_0 \gtrsim 20$ models
    (Figure~\ref{fig:v_sh}).

    In summary, the density ratio between the CSM and outermost
    ejecta for SN\,2014C can be constrained by radio, optical, and
    x-ray data to be $11 < A_0 < 750$ based only on the wall density
    (we discuss $\rhowall = 10^{-18}, 10^{-17}~\densu$)
    and time of impact (we discuss $\timp=100, 120, 190~\days$).
    These values of $A_0$ are all in the regime where 
    self-similar solutions do not accurately approximate the hydrodynamics.
    Radio VLBI measurements have been reported for SN\,2014C, 
    though we can only use the earliest of the observations for
    direct model comparison.
    The measured radius at 384 days 
    best matches models with $A_0\sim20$ for the earlier $\timp$
    and $A_0\sim60$ for $\timp=190~\days$,
    independent of  $s$, $\eta$, or $F_R$.
    For any $A_0>11$, 
    the measured deceleration parameter favors $s=2$ for the outer medium, 
    and strongly disfavors $s=0$, particularly if the wall was formed
    adiabatically such that $\eta=4$.
    If both $\eta=4$ and $s=2$, then for the baseline
    models ($F_R=1.1$), the measured shock velocity at 500 days 
    is most consistent with models that have $A_0\sim20$, but
    $3\lesssim A_0 \lesssim 190$ are within the velocity and $\timp$ uncertainties.
    We conclude that, analyzing the SN\,2014C VLBI 
    observations within the context of our models and assuming
    $\eta=4$, an $s=2$ outer medium is favored and 
    models with $\rhowall\sim10^{-18}~\densu$
    are consistent for the entire range of $\timp$, but the
    $\rhowall\sim10^{-17}~\densu$ wall proposed by M17 is also
    within errors as long as $\timp\sim100~\days$ rather than their
    assumed $\timp=190~\days$.
    Earlier impact times are also favored by the optical observations of
    interaction  signatures by 120 days, and 
    and imply $\Rsw\sim3\times10^{16}~\cm$,
    consistent with the x-ray non-detections of M17. 
    The wind profile derived by \citet{Tinyanont+19} 
    has a density $1.15\times10^{-18}~\densu$ at $2.6\times10^{16}~\cm$,
    in line with our range of $\eta$ and $\rhowall$.

   \subsection{A Reinterpretation of SN\,2014C}\label{ssec:14Crev}

    Both the x-ray and radio emission of SN\,2014C have maxima at
    $t\sim400-500~\days$ (A17, M17), and the x-ray emission measure
    indicates that the shocked CSM mass is $1-1.5~\Msun$ at this time;
    but what does this mean about the shock evolution?

    M17 interpret the x-ray peak (they use $t\sim500~\days$)
    to be the shock front passing over the CSM wall. 
    Maintaining this assumption but using
    the earlier impact time of $\timp=120~\days$ from optical observations, 
    a model with $A_0=50$ (in line with VLBI) has $M_\wall\approx1.2~\Msun$,
    approximately the derived mass of shocked CSM from the x-rays. 
    This wall would extend to $\Rwall \sim 6\times10^{16}~\cm$ 
    (similar to their derived $\Rwall$)
    but would have a lower density than they derived, 
    $\rhowall\sim2.7\times10^{-18}~\densu$.
    All wall models with $M_\wall=1-1.5$ and the above interaction timings
    are similar -- the allowed range is $A_0 \sim 30-70$, which sets 
    $F_R \sim 2-2.25$ and $\rhowall \sim (1-4)\times10^{-18}~\densu$.
    Thus our models suggest a lower-density, thicker wall compared to M17,
    because the mass is maintained but the inner radius decreases.

    Now we offer a more speculative extension of our hydrodynamic results
    that re-interprets the radio and x-ray peaks.
    Our re-interpretation stems from the fact that the shock speed determines
    the gas energy density and, for $A_0\gtrsim10$ values, 
    the shock speed peaks well after the shock has crossed the wall. 
    We speculate that the radio rise may actually reflect the rapid rise 
    of $\vshock$ \textit{following} the crossing of the wall, i.e., $\tx=190~\days$.

    The optically thin radio luminosity depends strongly on shock speed.    
    Using Equation~37 of \citet{HNK16}, the radio emissivity 
    ($j_\nu$, units $\mathrm{erg~s^{-1}~Hz^{-1}~cm^{-3}~sr^{-1}}$)
    is $j_\nu \propto u_\mathrm{gas}^3 \propto \vshock^6$,
    assuming the gas energy density ($u_\mathrm{gas}$) is 
    proportional to the square of the shock speed ($\vshock$).
    The luminosity will depend on this factor, the volume of shocked gas,
    and the optical depth, which all depend on the shock speed for their 
    time evolution, and should be dominated by changes in emissivity 
    (optically thin regime) and optical depth (optically thick regime).
    Unpublished data indicate that the 15.7~GHz light-curve was 
    optically thin at $\sim300~\days$ (A.~Kamble, priv.~comm.), 
    and, if the 15.7~GHz rise is optically thin, then it evolves like
    $v_\mathrm{sh}^6$.   
    The 15.7~GHz flux increased by a factor of $\sim6$ between 
    190 and 400 days (A17, Figure~1), requiring
    only a factor of $\sim1.35$ increase in $\vshock$.
    
    The magnitude and timescale of increase in $\vshock$ is matched
    by our models with a significantly lower wall mass than posited by M17.
    Allowing $\timp=100, 120~\days$, $\tx/\timp=1.9,1.6$ and 
    $t_\mathrm{peak}/\timp = 4, 3.3$.
    From Figure~\ref{fig:v_sh} (models have $F_R=1.1, \eta=4$, and $s=2$)
    we see that the $A_0\sim190$ model 
    (highest $A_0$ of the SN\,Ib set)
    has $\tx/\timp\sim2$, a peak speed at $t/\timp\sim4$, and
    an increase in shock speed of $\sim1.3$ between crossing
    and peak, which are all consistent with the 15.7~GHz light-curve
    under the optically thin assumption.
    The $A_0\sim60$ models peak at $t/\timp\sim3.25$ with a 
    shock speed increase of $\sim1.4$. 
    This reinterpretation implies 
    $F_R=1.25,1.15$ and $M_\wall=0.06,0.04 ~\Msun$ for $A_0=20$
    or $F_R=1.14,1.08$ and $M_\wall=0.31,0.17~\Msun$ for $A_0=200$
    (Equations~\ref{eqn:A0}, \ref{eqn:Rfwdwall}, \& \ref{eqn:Mwall}).
    At $t\sim500~\days$ the total mass of shocked CSM would 
    be $\sim1~\Msun$ (Figure~\ref{fig:M_sh}, $F_R=1.1$ models),
    in agreement with the shocked CSM mass estimate of M17, 
    but most of this mass is from the outer CSM, not the wall.
    
    Thus we find that if we assume an impact time $\timp \sim 100~\days$, 
    we can re-interpret the M17 derivation of a shocked CSM mass $1-1.5~\Msun$
    at 500 days in two ways.
    First, maintaining their assumption that $\tx\sim500~\days$, 
    we find the wall density must be $\sim2-10$ times lower than they
    report.
    However, we also posit $\tx\sim190~\days$, 
    and find this implies a wall
    $\sim5-40$ times less massive than they reported,
    and the mass at 500 days is primarily shocked wind material.
    In both cases, the range of values reflects uncertainty in $A_0$
    but implied $A_0$ values are in line with the constraints from
    VLBI.
    Radiation transport calculations are required for 
    calculations for detailed comparison to observations, 
    including accurate interpretation of the radio rise and peak of SN\,2014C,
    which we leave to our next publication on this model suite.

    \subsection{Memory of the Wall}\label{ssec:nowall}
    A wall of very limited extent may be crossed over by the
    forward shock very quickly, such that observations only probe
    the phase of evolution in the outer CSM or perhaps have a single
    epoch of observations in the wall phase.
    Or, for example, the hydrodynamics-probing observations 
    of \citet{Bietenholz+18} could only be undertaken at late times
    because the method requires the shock to have a certain 
    angular extent.
    Therefore, a crucial question is whether (and for how long) 
    the existence of a wall can be inferred from observations after 
    the wall-crossing time, i.e., how different is the evolution
    with a wall compared to with the outer CSM alone?
    
    As an initial probe of this question, 
    we removed the wall from the highest- and
    lowest-$A_0$ SN\,Ib simulations with $s=2$ and $\eta=4$ 
    (baseline values), extended the outer CSM inward to 
    maintain the impact time, and re-ran the simulation.
    Note that $A_0$ is reduced 
    by a factor of $\eta$ in a wind-only model compared to the
    with-wall version, but we will refer to $A_0$ values from the
    with-wall model.
    Because $F_R$ affects the initial conditions of the shock
    front evolution in the outer CSM, we compare the wind-only
    simulation to with-wall simulations of all $F_R$
    values (1.01, 1.03, 1.1, and 1.3). 
    We will be comparing the difference in hydrodynamic quantities
    between the no-wall models and their counterparts with a wall.
    We will compare the simulation differences
    to observational errors from VLBI
    of SN\,1993J at late times \citet{Bartel+02} and
    SN\,2014C that are likely at $t/\timp < 10$ according
    to our analysis \citep{Bietenholz+18}.
    
\begin{figure}
        \centering
        \includegraphics[width=0.7\linewidth]{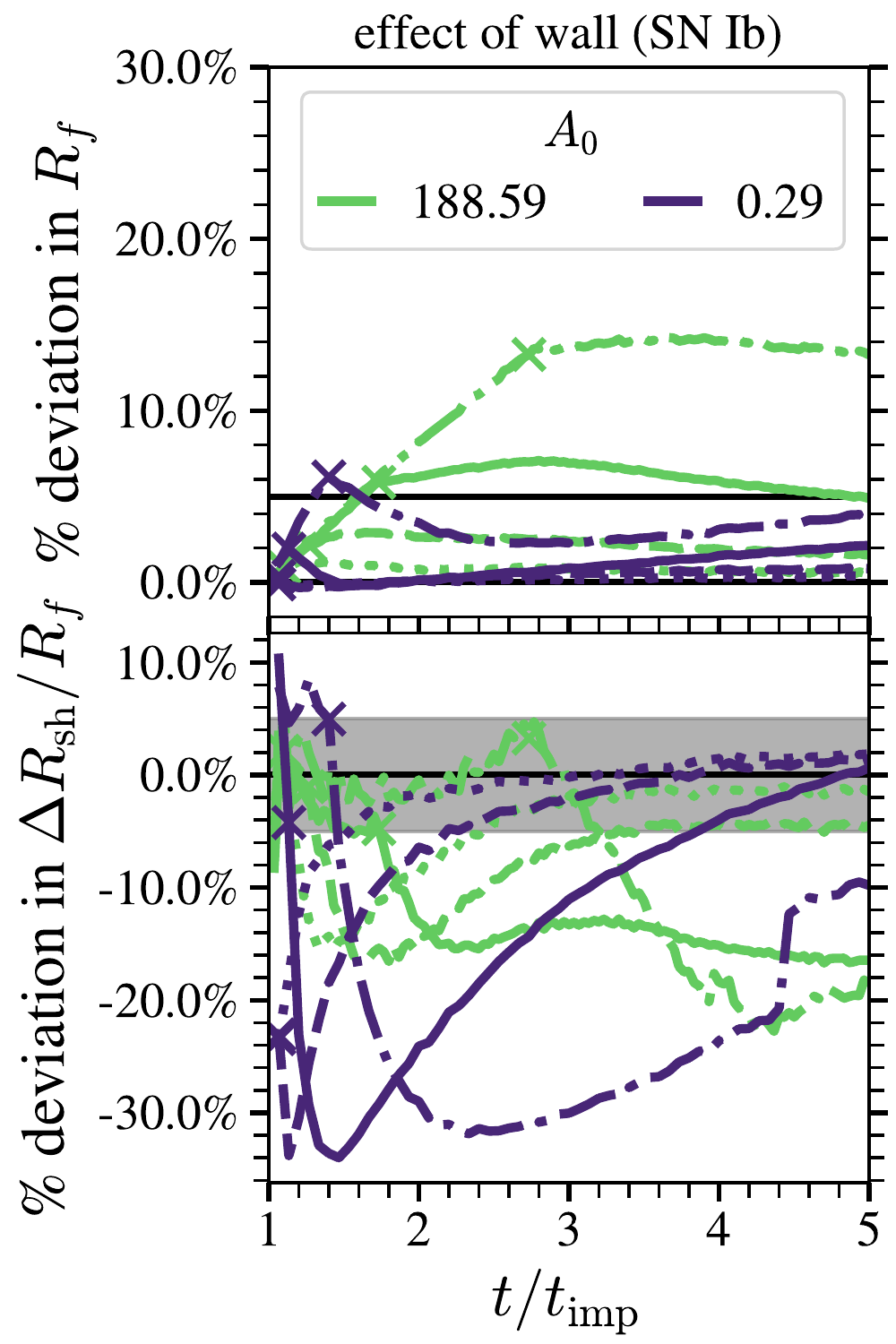}
        \caption{Comparison of model without the wall to the suite 
        model with a wall for SN\,Ib. 
        Line style shows $F_R$ as in Figure~\ref{fig:r_sh_svar_plaw}.
        Color denotes $A_0$ (see legend) amd
        an ``x'' marker shows $\tx$.
        \textit{Top.} Forward shock radius deviation,
        $(R_{f,\mathrm{no-wall}} - R_f)/R_f$.
        \textit{Bottom.} Shock thickness, $(y_\mathrm{no-wall} - y)/y$,
        where $y = \Delta R_\mathrm{sh}/R_f$.
        }
        \label{fig:nowall}
    \end{figure}

    Figure~\ref{fig:nowall} shows the comparison of no-wall
    models to models in the suite with the same outer CSM.
    Line style represents $F_R$ as in  Figure~\ref{fig:r_sh_svar_plaw},
    color represents $A_0$ as given in the legend and the same as
    in all other plots.
    The top panel shows the deviation in forward shock radius
    of the model without a wall from the model with a wall
    (black lines are 0\% and 5\%),
    and the bottom panel shows the same for the shock width
    (black line is 0\%, grey band shows $\pm5\%$).
    
    We find the percentage deviation between wall/no-wall models 
    in forward shock radius (top panel) is small for
    most models.
    It is often comparable to observational error, which
    for SN\,1993J was $<1\%$ and for SN\,2014C  
    $3\%-23\%$ (typically $\sim4\%$).
    We also looked at the deceleration parameter ($m$) that would
    be measured in $3 \leq t/\timp \leq 5$.
    We found that $m$ fit to the model without a wall deviated 
    from the wall model by $(0.5-5)\%$, depending on $F_R$ and $A_0$.
    The errors on $m$ reported for SN\,1993J were $(1-3)\%$
    and for SN\,2014C, $\sim5\%$. 
    Therefore it does not seem the measurements of $m$ are
    precise enough to say whether a wall is present -- 
    and according to our analysis, interpretation of $m$
    is complicated anyway from other CSM parameters
    (\S~\ref{ssec:shock_r}, Figure~\ref{fig:r_sh_svar_plaw}).

    The width of the shock region (bottom panel)
    does seem to be a potential probe of the presence of a wall.
    At most times, the shock region is thicker
    when a wall is present. 
    This is in part because the forward shock radius, which
    we use to normalize the thicknesses, is higher in wind-only
    models.
    The other contribution is from the wall driving 
    a stronger shock back into the ejecta, 
    widening the shock region.
    The sudden change seen in the low-$A_0$ $F_R=1.3$ model 
    curve at$t/\timp \sim 4.5$ is due to the shock front finder
    identifying a rarefaction wave, 
    also seen in Figure~\ref{fig:thickness},
    as described in \S~\ref{ssec:an_meth}.
    Grey bands show $5\%$ deviations, which is comparable
    to the precision of shock thickness measurements for 
    SN\,1993J, albeit at late times.

    We conclude that for kinematic quantities about the shock, 
    e.g., those probed by VLBI, observations would probably be
    equally well modeled by a cavity and wind as 
    by a cavity, wall, and wind, with the wind having
    the same properties between both models.
    If the shock thickness can be measured, an anomalously
    thick shock region may indicate that the CSM had a wall.
    
    However, other radiation signatures will likely be very 
    different with or without a wall -- 
    \citet{Dwarkadas+10} noted that a wall
    was required to match the x-ray evolution of SN\,1996cr,
    and in our own exploratory modeling of SN\,2014C we found
    this to be the case as well.

\section{Summary}\label{sec:conc}

We explore the impact of a normal
SN\,Ia or SN\,Ib with a circumstellar medium (CSM)
that has been shaped by an eruption or change in wind 
properties --
a smooth distribution of outlying mass is partially
swept up into a ``wall'' of material at $\sim10^{16}~\cm$.
Our interest is constraining the CSM of canonical events 
like PTF\,11kx or SN\,2014C and providing a standard 
baseline for interpreting future events and observational
ensembles.
Through a suite of  $\sim600$ one-dimensional models, we 
traverse a wider range of parameter space in CSM properties
than any similar study yet undertaken.
An overview of the parameters in this study is illustrated in
Figure~\ref{fig:init_conds}.
Our baseline values are $s=2$ (outer medium is a wind), 
$\eta=4$ (the wall is formed by adiabatic compression of the wind
in a strong shock), and $F_R=1.1$ (10\% fractional extent of the wall). 
By running these simulations with \PaulsCode\ we are able to
study the effects of mixing due to the Rayleigh-Taylor
instability at the interface of the shocked media, a well-known
yet rarely captured effect in studies of supernova-CSM interaction.
Since it is nevertheless a one-dimensional model, it does not 
capture the two-phase nature of the turbulent area, 
e.g., to create dense clumps.
Details of our simulations can be found in \S\,\ref{sec:models}.
Our primary assumptions are that the SN ejecta can be described
by a broken power law, that the maximum ejecta velocity is 
$30,000~\kms$, that the hydrodynamics can be described by an 
adiabatic index $\gamad=5/3$ throughout the evolution,
that the wall is a constant density, and that the 
CSM interior to the wall is low enough density that it can be
ignored.

In this manuscript, we limit our analysis of the simulation suite
to hydrodynamic properties of the shock:
the evolution of the forward shock front radius,
evolution of shock front speeds,
deceleration of the ejecta, 
mass of shocked material, 
and amount of mixing between the shocked ejecta and CSM.
Details of our analysis methods can be found in \S\,\ref{ssec:an_meth}.
Our main conclusions are as follows.
\begin{enumerate}
\item The initial ratio of the CSM density to the density of outer ejecta, $A_0$
      governs the evolution of the shock (\S~\ref{sec:analysis}). 
      This is in line with \citet{Dwarkadas05}.
\item When the reverse shock reaches ejecta of similar density to $\rhowall$
      the self-similar solution applies for describing hydrodynamic properties
      (\S~\ref{ssec:shock_v} \& \ref{ssec:ej_v}).
      This occurs later for higher-$A_0$ models.
\item We find a simple function for forward shock radius while inside the wall
      with parameters that depend on $A_0$ (\S~\ref{ssec:shock_r}, 
      Figures~\ref{fig:r_sh}~\&~\ref{fig:r_sh_plaw}).
\item At late times, the shock radius evolves
      as a power law, so a deceleration parameter ($m$) can be measured
      (\S~\ref{ssec:shock_r}, Figure~\ref{fig:r_sh_svar}).
      We find that $m$ only indicates the CSM density profile $s$ 
      if $A_0$ is very low; 
      for higher values of $A_0$, $\eta$ and $F_R$ change $m$ 
      as much as $s$ does
      (Figure~\ref{fig:r_sh_svar_plaw}).
\item The thickness of the shocked gas grows to $\sim 20\%$ of the forward 
      shock radius by $\sim 2\timp$, independent of $A_0$, CSM configuration, 
      or SN type. 
      However, at early times, it grows rapidly, which should be taken into
      account when estimating the volume of shocked gas 
      (\S~\ref{ssec:shock_dr}, Figure~\ref{fig:thickness}).
\item The reverse shock traverses the ejecta faster for a higher $A_0$, 
    so the self-similar solution breaks down more quickly by reaching
    the inner ejecta for higher-$A_0$ models
    (Figures~\ref{fig:mod_ev}, \ref{fig:vej}, \ref{fig:M_sh}).
     Taken together with the point 2 above, the self-similar solution
     has a much more limited time frame of applicability in high-$A_0$
     situations.
\item The deceleration of the ejecta is significantly greater from higher
    walls than would be calculated from the self-similar solution 
    (\S~\ref{ssec:ej_v}, Figure~\ref{fig:vej}).
    Observations of delayed-interaction SNe may indicate 
    $A_0>100$ is common.
\item The fraction of shocked ejecta within the total shocked material
    varies with $A_0$ and time.
    Especially at early times, it cannot be assumed that the same mass of 
    ejecta has been shocked as CSM -- for high $A_0$ it may be only 10\%
    (\S~\ref{ssec:shock_m}, Figure~\ref{fig:M_sh}).
\item The fraction of shocked CSM that has ejecta mixed into it 
    varies with $A_0$ and time (\S~\ref{ssec:mixing}, Figure~\ref{fig:mixed_frac}).
    Generally, at least 10\% of shocked CSM is polluted by ejecta at the
    $\geq 1\%$ level.
\item Applying our models to SN\,2014C, under the assumption $\eta=4$
      we find the VLBI observations agree most with a wall having
      $11\lesssim A_0 \lesssim200$ and outer CSM of $s=2$ 
      (\S~\ref{ssec:14C}). 
\item We suggest that radio rise of SN\,2014C is due to rapid 
      shock acceleration after wall crossing, in which case the wall 
      mass is only $M_\wall=0.04-0.31~\Msun$, much lower than derived by
      M17 (\S~\ref{ssec:14Crev}).
\item We tested the effect of the wall on late-time hydrodynamics for a 
      few models, and find that the difference in shock radius and radius
      time evolution are comparable to observational error. 
      However, the shock may be measurably wider even at late times
      when a wall is present (\S~\ref{ssec:nowall}).

\end{enumerate}

The most directly applicable observations for comparing to our results 
are those of very long baseline interferometry (VLBI), 
which unfortunately requires that the SN be very nearby. 
Optical line profiles have in the past been used to infer the
bulk gas speed of shocked material, which can be compared to 
the shocked gas speeds presented here.
In a sequel paper we will compute continuum radiation for this
model suite, which will enable a wider array of quantiative 
comparisons to observation.
We anticipate that the insight gained from the hydrodynamic behaviors 
presented here will be useful when interpreting the features of those 
light-curves.


\section*{Acknowledgements}
The authors thank Paul Duffell for assistance with
using \PaulsCode\ and helpful discussions.
The authors would like to acknowledge
Raffaella Margutti, Dan Milisavljevic, Daniel Kasen, 
Ken Shen, Laura Chomiuk, Sumit Sabadhicary, and Sean Couch
for helpful discussions during the preparation of the manuscript, 
and the helpful comments of the anonymous reviewer.
Simulations were run on Sparky, a workstation 
funded by the Department of Energy Computational Science Graduate Fellowship (CEH).

CEH acknowledges the Anishinaabek as 
the caretakers of the land on which she undertakes her work.

\software{
RT1D \citep{Duffell16},
SciPy \citep{SciPy}, 
NumpPy \citep{NumPy}, 
Astropy \citep{Astropy},
Matplotlib \citep{Matplotlib}
}
\bibliographystyle{apj}
\bibliography{refs}

\end{document}